\documentclass[
&aps,
prl,
&prd,
twocolumn,
&onecolumn
]{revtex4-2}

\newcommand{\beqa}{\begin {eqnarray}}
\newcommand{\eeqa}{\end {eqnarray}}

\usepackage{float}
\usepackage{natbib}
\usepackage{amsbsy}
\usepackage{amssymb}
\usepackage{amsmath}
\usepackage{graphicx}
\usepackage{mathrsfs}
\usepackage[caption = false]{subfig}
\usepackage{array}
\usepackage{mathrsfs}% for script P
\usepackage{xfrac}
\usepackage{booktabs}
\usepackage{amsthm}
\usepackage{hyperref}
\hypersetup{colorlinks=true, citecolor=blue, urlcolor=blue, linkcolor=blue}

\usepackage{mathtools} % Need for cases.
\usepackage{graphicx}% Include figure files
\usepackage{dcolumn}% Align table columns on decimal point
\usepackage{bm}% bold math
\usepackage{etoolbox}
\apptocmd{\sloppy}{\hbadness 10000\relax}{}{}

\usepackage{xcolor}
\usepackage{amsthm}
\usepackage{pifont}

\usepackage [english]{babel}
\usepackage [autostyle, english = american]{csquotes}
\usepackage{natbib}
\MakeOuterQuote{"}
\usepackage{enumerate}

\def\be{\begin{equation}}
\def\ee{\end{equation}}
\def\v{\cal{V}}

\begin{document}
\title{ Disentangling the growth rate of perturbations from the HI bias  using only clustering data from galaxy surveys }% Force line breaks with \\
\author{Pankaj Chavan}\email[E-mail: ]{chavanpankaj09@gmail.com}
\author{Tapomoy Guha Sarkar}
\email[E-mail: ]{tapomoy1@gmail.com}
\affiliation{Department of Physics, Birla Institute of Technology and Science - Pilani, Rajasthan, India}
\author{Anjan A Sen}
\email[E-mail: ]{aasen@jmi.ac.in}
\affiliation{Centre for Theoretical Physics, Jamia Millia Islamia, New Delhi-110025, India}

\begin{abstract}
This work serves two-fold purpose. Firstly, we provide an alternative to the traditional method of determining the growth rate of density perturbations $f(z)$. In usual practice, $f(z)$ can not be directly measured from tracer clustering at some redshift without knowledge of the bias. While the bayron acoustic oscillation (BAO)  imprint allows the determination of $(D_A(z), H(z))$, redshift space anisotropy (RSD) allows the measurement of a quantity $f_8(z) = f(z) \sigma_{8,0} D_{+}(z)$. To extract $f(z)$ from $f_8(z)$, one usually requires some other data set. We show that precise BAO and RSD measurements in and around some key redshifts themselves can solely reconstruct $f(z)$ without requiring any other data sets. 
Secondly, we extend this approach to another tracer, namely the post-reionization 21-cm brightness temperature intensity maps. We demonstrate that the measured $f(z)$ from purely redshift space clustering allows us to measure the 21-cm bias, which is a largely unknown quantity. This may help interpret the observed intensity mapping signal in the future.

\end{abstract}

\maketitle

{\textit{Introduction:}} One of the central goals of observational cosmology is to understand how small initial matter density perturbations \cite{sachs1986perturbations}, around a smooth background, grew under gravitational instability in an expanding Universe to form the rich structure of the cosmic web \cite{bond-cosmicweb, White-cosmicweb}. 
 Assuming that dark energy does not cluster \cite{Batista_clustering-de}, cosmological structure formation is primarily driven by dark matter. A key quantity that encapsulates the growth of matter perturbation is the \textit{linear growth rate}, $f(z) \equiv - (1+z)  { d {\rm ln} D_+(z)}/{d z}, $ where, $D_{+}(z)$ is the growing solution for the temporal part of the dark matter overdensity field. The sensitivity of this growth rate to cosmological models makes it an important probe for distinguishing between $\Lambda$CDM and alternative models \cite{Maeder_2017_LCDM_alternative, BULL_et_al_2016_Beyond_LCDM, Arun_2017_DM_DE_alternative_models}. 
Dark matter distribution and clustering are probed using indirect tracers like galaxy surveys \cite{Hou_2020, Beutler_2012_6dF_growthRate, Alam_2017_SDSS_III, Howlett_2015_fs8, Percival_2007}, Lyman-$\alpha$ forest, and 21-cm intensity maps \cite{mcdonald2006lyalpha, irvsivc2017lyman, Slosar_2011, Garzilli_2019}. 
 Measuring $f(z)$ directly from the clustering of tracer fields is challenging. Instead, a related quantity: $f_8(z) \equiv f(z) \sigma_{8,0} D_{+} (z)$, is a key observable in galaxy redshift surveys. It is directly constrained by redshift-space distortions (RSD), where it imprints itself due to peculiar flow, and manifests as an anisotropy in the redshift space clustering. In linear theory, this effect is described by the Kaiser formula \cite{hamilton1998linear}, which modifies the real-space galaxy power spectrum $P^{G}_{real}(k)$ to give the redshift-space power spectrum $P^{G}_s(k, \mu)$:
$ P^{G}_s(k, \mu) = \left(1 + \beta \mu^2 \right)^2 P^{G}_{real}(k),$
where $\mu$ is the cosine of the angle between the wavevector $\vec{k}$ and the line of sight, and $\beta_G = f / b_G$, with $b_G$ being the linear galaxy bias. 
 The amplitude of the galaxy power spectrum is proportional to $(b_G\sigma_8)^2$, and the RSD anisotropy depends on the parameter  $\beta_G = f/b_G$. The combination $
f_8(z) = \beta_G \cdot (b_G\sigma_8)$
is thus directly accessible from observations and is measured more robustly than $f(z)$ or $\sigma_8(z)$ individually. 
Galaxy surveys such as BOSS \cite{Delubac_2015, FontRibera_2012}, eBOSS \cite{Zhao_2016_eBOSS, Ivanov_2021_elg}, and DESI \cite{DESI_Colab_2025_DR2, Adame_DESI_Data_2025, levi2013desi} have measured $f_8(z)$ across a wide range of redshifts with increasing precision. Moreover,  surveys like Euclid \cite{Scaramella_2022_euclid_prepI} aim to achieve sub-percent level precision on $f_8$, providing even stronger tests of cosmology. Traditionally, measurement of $f(z)$ would either require knowledge of $\sigma_8(z)$ (from CMBR) \cite{PLANCK18_COSMO_params_Aghanim_2020} or by combining RSD with weak lensing \cite{abbott2022_DESY_weaklensing, Porredon_DESY_lensing_2022, Pandey_DESY_gal_gal_lensing_2022, Hikage_2019_CosmicShear_PS, Karim_2025_sigma8} or velocity data \cite{Nusser_2012_log_grth_rt}. 

Although galaxies have been the most useful tracer candidate, the post-reionization H\,\textsc{i} 21 cm brightness temperature maps are believed to be a very promising tracer of the underlying dark matter distribution \cite{poreion0, poreion1, poreion2, poreion3, poreion4, poreion5, poreion6, poreion7, poreion8, poreion9, poreion10, poreion10, poreion11, poreion12}. This makes tomographic intensity mapping of the post-reionization H\,\textsc{i} 21-cm signal a potentially powerful observational probe of background cosmological evolution and structure formation. 

However, the distribution of neutral hydrogen (H\,\textsc{i}) in the post-reionization Universe remains unknown. Baryonic matter is expected to follow the underlying dark matter distribution on large scales with a possible linear bias; however, this assumption fails at low redshifts and small scales due to nonlinear gravitational clustering. It is, however expected that H\,\textsc{i} traces the matter distribution with a scale- and redshift-dependent bias $ b_T(k, z) = \left[ {P_{ H\,\textsc{i}}(k, z)}/ {P_m(k, z)} \right]^{1/2}$
where, $P_{ H\,\textsc{i}}$ and  $P_m$ are the  H\,\textsc{i} and dark matter power spectra, respectively. 
It is imperative to model this bias accurately for any cosmological interpretation of the  21-cm intensity mapping observations. The standard method to model the bias requires N-body simulations \cite{Bagla_2010, Guha_Sarkar_2012, Sarkar_2016}. In these simulations, dark matter halos are identified, and then they are populated with  H\,\textsc{i} according to empirical rules. Following the completion of reionization, H\,\textsc{i} primarily resides in damped Lyman systems, and the collective diffuse emission from these sources is believed to trace the underlying dark matter distribution. Assuming a constant neutral fraction \cite{xhibar1, xhibar2}
 $\bar{x}_{ H\,\textsc{i}} = 2.45 \times 10^{-3}$, the power spectrum of post-reionization H\,\textsc{i} 21-cm excess brightness
temperature field $\delta T_b$ from redshift $z$  is given by \cite{mcquinn2006cosmological, Bull_2015}
\be 
P_T ( k, z, \mu ) = A_T(z)^2  \left (b_T(k,z)  +  f(z) \mu^2 \right ) ^2 P_m(k,z)\ee 
where
\be
A_T = 
4 {\rm mK} \bar{x_{HI}} ( 1 + z)^2 \left ( \frac {\Omega_n h^2}{0.02} \right) \left ( \frac{0.7}{h} \right ) \left ( \frac{H_0}{H(z)} \right).
\ee
Observation of the 21-cm power spectrum can allow a measurement of the distortion parameter $\beta_T = f/b_T$.
To measure the $b_T$, one has to measure  $f(z)$ independently.
We have discussed earlier that $f(z)$ is not directly measured using RSD in galaxy surveys. 
\begin{figure}[ht]
\centering
\includegraphics[height=5.5cm, width=8.5cm]{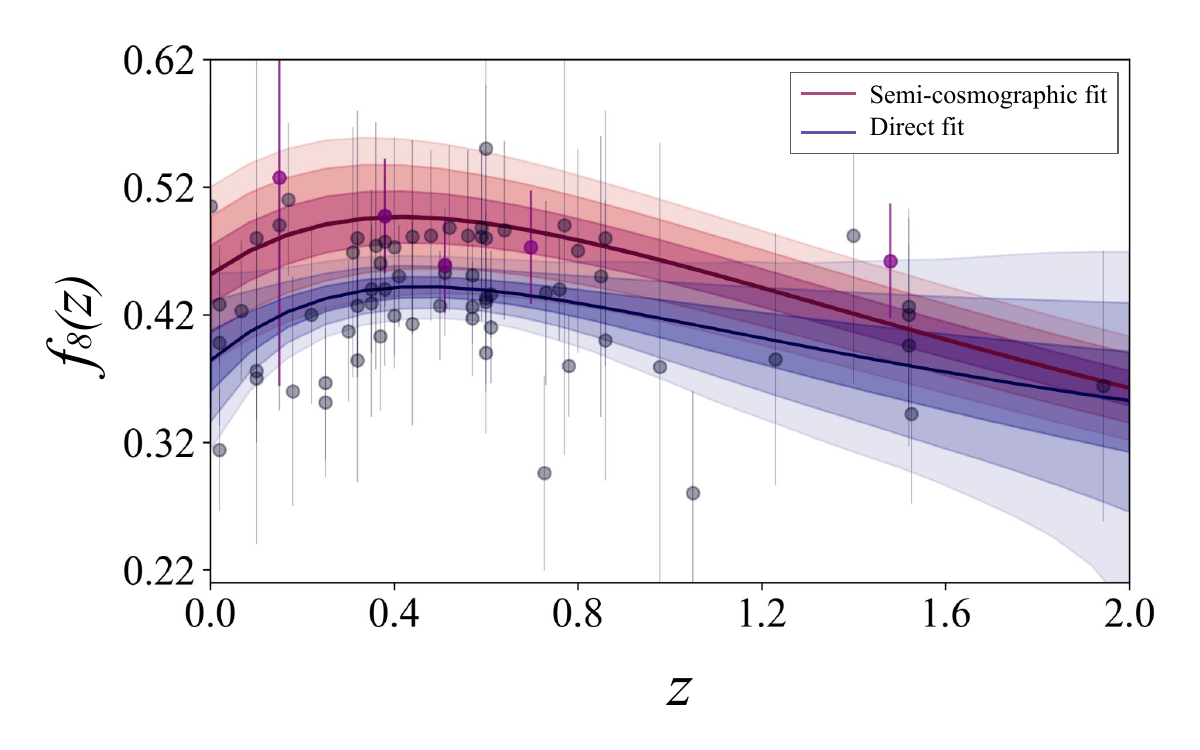}
%\captionsetup{skip=0.3pt}
\vspace{-15pt}
\caption{shows the the reconstruction of $f_8(z)$  by two entirely different approaches. In the semi-cosmographic fit, the $(H, D_A, f_8)$ data from SDSS IV is used to fit model parameters, while in the other case, all available $f_8$ data is directly fitted with a fitting function.}
\label{fig:feightrecon}
\end{figure}

In this work, we propose a method by which $f(z)$ can be measured at some specific fixed redshifts,  only using galaxy clustering data. We show that using the BAO and RSD data on $H(z) $ and $f_8(z)$  from galaxy surveys can allow us to measure $f(z)$ at a few fixed redshifts without requiring CMBR or weak lensing data. 
Since 21-cm maps are tomographic,  one may tune the frequency of radio observation to frequencies corresponding to these specific redshifts and thus measure $b_T$ on large scales from a measurement of $\beta_T$ at these redshifts. Central to our method is the determination of $f(z)$  from $f_8(z)$  and $H(z)$ data only in galaxy surveys. We now demonstrate that this is possible when precise data on $f_8(z)$ and $H(z)$ is available.

{ \textit {Formalism: }}
  By using the sound horizon $r_d$  at the drag epoch as a standard ruler and by using the  Baryon Acoustic Oscillations (BAO) feature in the transverse and radial clustering of Dark matter tracers (say galaxies), it is possible to simultaneously measure the Hubble expansion rate $H(z)$ and the angular diameter distance $D_A(z)$ at a certain redshift. 
Measurement of the anisotropic galaxy power spectrum in redshift space also allows us to measure $f_8(z)$. The dynamics of  background cosmology and structure formation can be seen as a trajectory in the phase space of the quantities  
$x=(H_0/c)D_A(z)$, $p = dx/dz$, $d_+  = \sigma_{8,0} D_+$ and $f_8(z) = f(z) \sigma_{8,0} D_+(z)$.
where $x$ and $p$ are related in a spatially flat cosmology through $ x + p(1 + z ) = H_0/H$.  In a cosmology where dark energy is characterized by an equation of state $w(z)$, the  triplet of directly observable quantities  $(x, p, f_8)$ is obtained from the dynamical system \cite{P_chavan_2025_semiCosmography_xpf8}
\begin{eqnarray}
& x'& = p  \nonumber \\
&{p}'& =-\frac{2p}{1+z} - \frac{3}{2E} \frac{1 + w }{~(1+z)^2} \nonumber   + \frac{3}{2E^3} \Omega_{m0} w(1 + z) \nonumber \\
 &d_{+}'& = -\frac{f_{8}}{1+z} \nonumber \\
    &f_{8}'& =\frac{3}{2} \frac{\Omega_{m0}(1+z)^2} {E^2}\left(wf_{8}-d_{+}\right) + \frac{f_{8}}{2}\left(\frac{1-3w}{1+z}\right)
\label{eq:autonomous_bgrnd_evln}
\end{eqnarray}
where $E^{-1} = x  + p(1+z)$. The first two equations can be solved independently of the latter two, with initial conditions $(x(z=0) = 0, ~p(z=0) =1)$. The solution of the equations for  $d+$ and $f_8$  requires the solution of the first two equations and initial conditions $d_+(z_{in}) = \frac{\sigma_8}{1+z_{in}}$ and $f_8(z_{in})  = \frac{\sigma_8}{1+z_{in}}$, at some early epoch $z_{in}$ in the matter-dominated epoch.
For any cosmological model with parameters $(H_0, \Omega_{m0}, \sigma_{8,0}, w(z))$ the entire dynamical evolution may be studied in the $(x, p, f_8)$ phase space .
The functions $x$ and $f_8$ exhibit peaks at some redshifts, respectively. Thus, the phase trajectories remain bounded in the phase space. This is because $D_A = {r(z)}{/( 1 +z)}  $ and $f_8 \sim f(z) / (1+z)$ (since $D_+(z) \sim \frac{1}{1+z}$ beyond some redshift) both involve a growing function in the product with monotonically decreasing $1/(1+z)$.
The behavior of $f_8$ near its peak is key for this work. We define a logarithmic slope  \be  F_1 \equiv \frac{d \ln f_8}{ d \ln a}   = \frac{3}{2} \left [ R \left ( \xi - w \right ) + \left (   w - \frac{1}{3} \right)  \right ]
\ee
where $\xi(z) = f^{-1}(z)$ and $R$ is the fractional contribution of dark matter to the expansion rate $R(z) = \Omega_{m0} (1+z)^3/ E^2(z)$. If the  peak of $f_8(z)$ occurs at a certain redshift $z_1$, then $F_1(z_1) = 0$ and 
\be \implies f(z_1)^{-1} =  w(z_1)  +  { \left ( w(z_1) - \frac{1}{3} \right) }{R(z_1)}^{-1}.
\ee
Thus, locating the peak redshift $z_1$ allows us to evaluate $f(z)$ at $z= z_1$, by computing the background quantities $E(z_1)$ and $w(z_1)$ at $z=z_1$. The reconstruction of these background quantities is possible from the BAO data on $D_A(z)$ and $H(z)$. 
For example, 
\be
    w (z) =\frac{2(1+z) {H}' (z)}{3 H(z)( 1 - R(z))} - \frac{1}{R(z)}.
    \label{eq:eos}
\ee
Thus, if galaxy clustering allows us to measure $f_8(z)$, $D_A(z)$ and $H(z)$ at several redshifts around $z_1$ where $f_8(z)$ peaks (or in other words $F_1$ vanishes), then $f(z_1)$ can be obtained from a knowledge of $(\Omega_m, E(z_1), w(z_1))$ without requiring any knowledge about the bias or $\sigma_8$. Thus, CMBR or weak lensing data won't be required to estimate $f$ at $z_1$. All this would require is a very good dataset of $f_8$
and $H(z)$ around $z_1$, such that the derivatives can be reconstructed.

If we now consider the quantity 
\begin{eqnarray}
F_2 &\equiv& \frac{d \ln F_1}{d \ln a} + F_1 \nonumber \\
F_2(z)  &=&\frac{3R}{2} \left [ 1 + w'(1+z) \frac{R-1}{R}+ 3w(R-1) ( w - \xi ) \right ]  \nonumber\\
&+&  \frac{3}{2} \left ( 1 - \frac{3}{2} R\xi \right )    \left [ R \left ( \xi - w \right ) + \left (   w - \frac{1}{3} \right)  \right ] \end{eqnarray}
If $F_2$ has a zero at $z=z_2$ then $f=\xi^{-1}$ can be obtained at $z=z_2$ by solving the equation $F_2(\xi, z_2) =0$ with $R = R(z_2)$, $w= w(z_2)$ and $w' = w'(z_2) $.
Thus, if $f_8(z)$ is well constrained from data, especially around the redshifts where its derivatives vanish, then one may well constrain a series of functions involving $f_8$ and its derivatives like 
\begin{eqnarray}
F_0&=&f_8, ~~ F_1 = \frac{d \ln f_8}{ d \ln a}, ~~~  F_2 = \frac{d^2\ln f_8}{ d (\ln a) ^2} + F_1 , \nonumber \\
F_3 &=&  \frac{d^3\ln f_8}{ d (\ln a) ^3} F_1 + \left ( \frac{d^2\ln f_8}{ d (\ln a)^2} \right )^2,... ~etc. 
\end{eqnarray}
If these functions can be reconstructed from measured $f_8$, then at redshifts corresponding their zeros,  $\{ z_m: F_m (z_m) = 0 \}$, 
we can find  $f(z_m)$ by solving a polynomial equation in $\xi = 1/f$ of the form $ 
F_m(\xi; z_m,  R , R', R'' ... w, w',  w''...) = 0 $, 
where, the background quantities $(R, R', R''...w, w', w''..)$ are  to be evaluated at $z_m$.
\begin{figure}[hb]
\includegraphics[height=5.5cm, width=8.5cm]{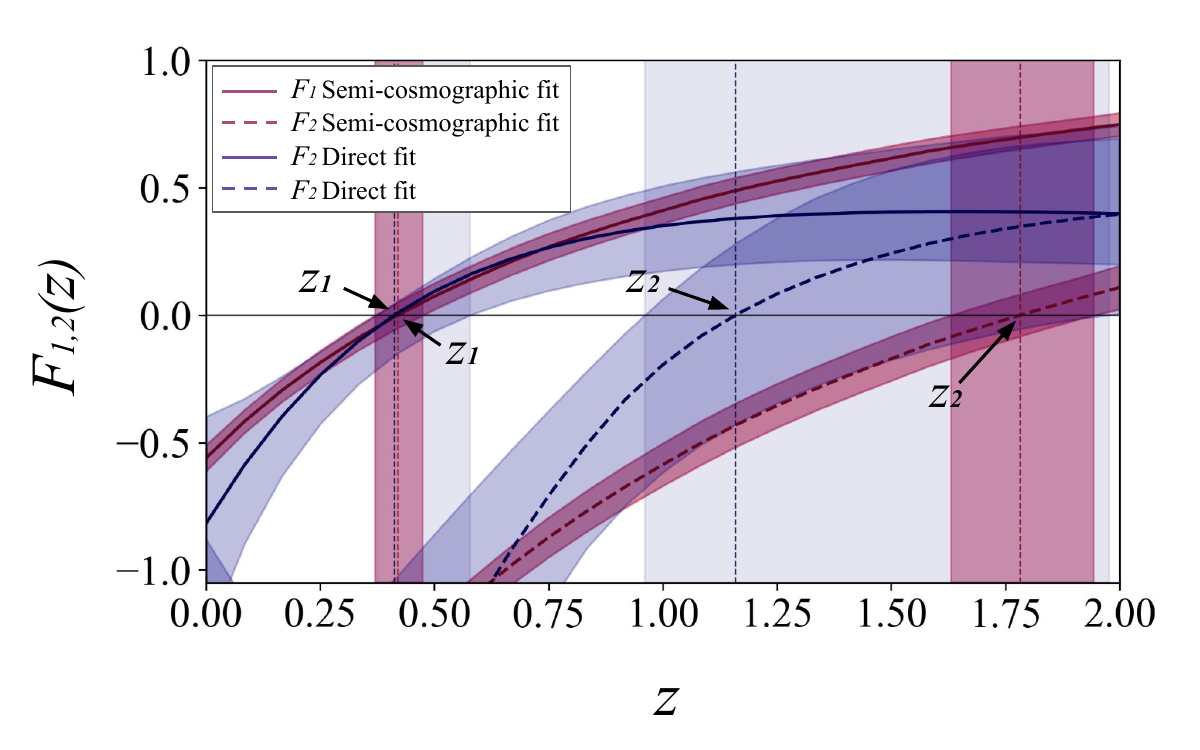}
\vspace{-15pt}
\caption{shows functions $F_1(z)$ and $F_2(z)$ reconstructed using both semi-cosmographic  and direct fitting method. The zero crossings $z_1$ and $z_2$ of these functions are marked respectively. While the slope reconstruction allows a better identification of $z_1$, there are large errors in $z_2$  due to a poor reconstruction of the double derivative.}
\label{Fig: zeros_F1_F2_vs_z}
\end{figure}
The method crucially requires a precise knowledge of the redshifts where $F_m(z)$ vanishes. The overall magnitude of $f_8$ has no bearing on the shape and therefore on its analytic properties.
In our method, the data on $f_8(z)$ is only used to find the redshifts where the function and its derivatives change direction. To determine $f$ at these redshifts, we have to evaluate some background quantities which are usually well constrained from the same or other data sets at these redshifts. 

In this work, we demonstrate this method by obtaining $f(z)$ at two redshifts $z_1$ and $z_2$ by employing the functions $F_1$ and $F_2$ reconstructed from $f_8$ data and the background quantities from BAO data. The reconstruction of $f_8$ is needed to get the redshifts ($z_1, z_2$), and the reconstruction of the background evolution is required to subsequently find $f(z_1)$, and $f(z_2)$. 

We apply this idea to the post-reionization 21-cm signal.
The redshifted 21-cm signal from the post-reionization epoch is modeled using simulations. The bias $b_T$ adopted for most analytic work can not be directly measured unless one adopts a model-dependent value of $f(z)$. However, if the RSD  parameter $\beta_T$ is measured, then from the knowledge of $f(z)$ at some specific redshifts (using our proposed method), one shall have an observational constraint on the bias. This shall greatly improve our understanding of the H\,\textsc{i} distribution on large scales at those epochs and thereby improve the simulations which are used to predict the signal.

\begin{figure}[hb]
\includegraphics[height=5.5cm, width=8.5cm]{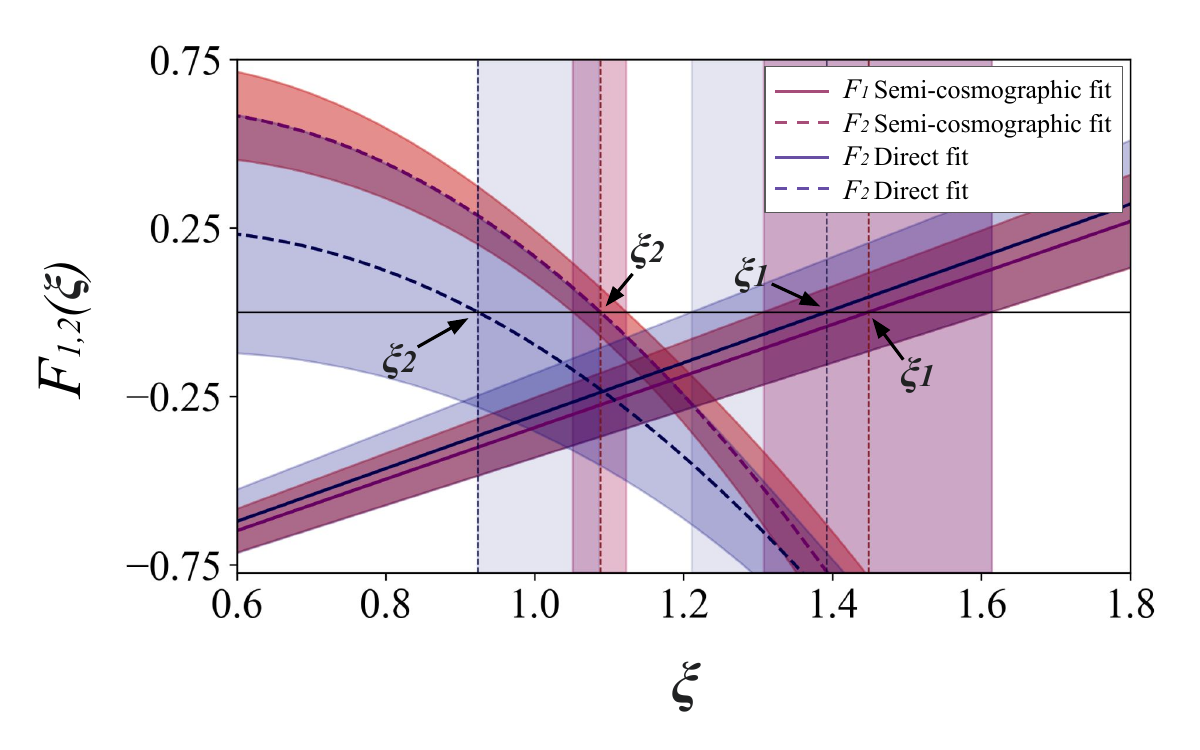}
\vspace{-15pt}
\caption{Demonstrating  the graphical method to find roots of $F_1(\xi_1) =0$ and $F_2(\xi_2) =0$, where all the background quantities appearing in $F_1$ and $F_2$ are evaluated at $z_1$ and $z_2$ respectively. }
\label{fig:zeros_F1_F2_vs_xi}
\end{figure}
\begin{figure}
    \includegraphics[height=5.5cm, width=8.5cm]{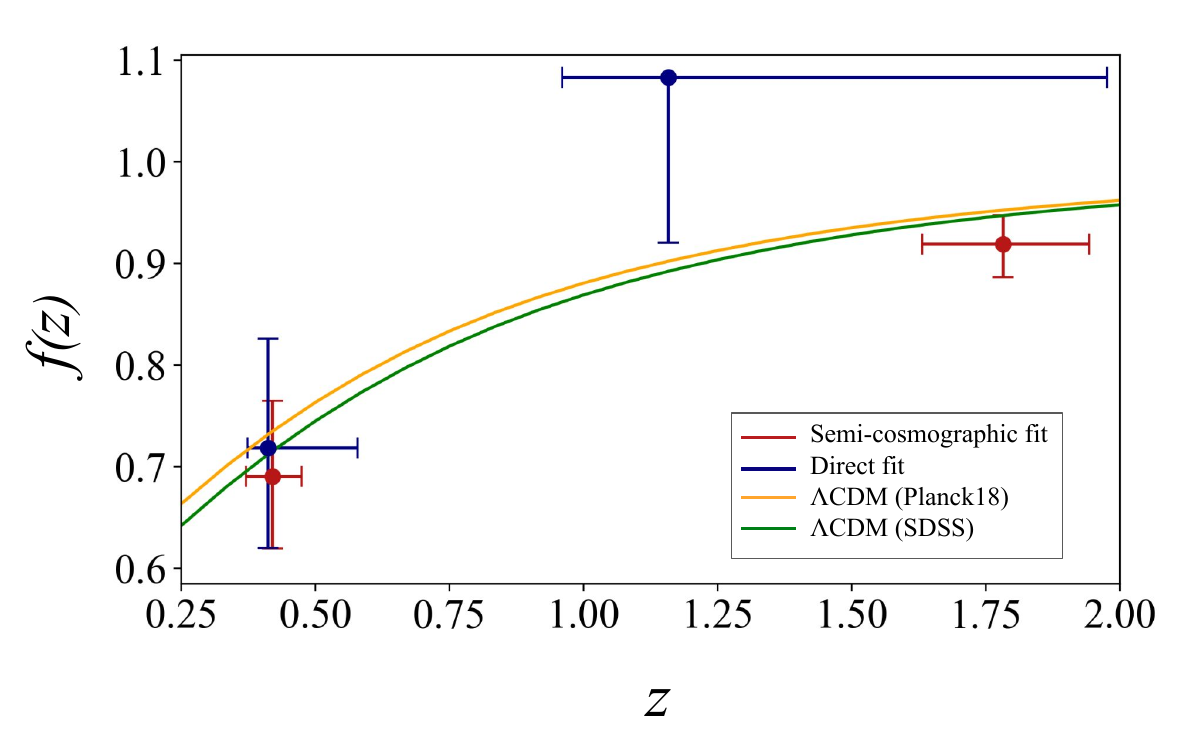}
    \vspace{-15pt}
    \caption{The reconstruction of $f(z)$ at two redshifts with $1\sigma$ error bars. The theoretical behaviour of $f(z)$ for the $\Lambda$CDM model with SDSS IV and Planck 18 fit parameters is also shown. }
    \label{fig:errorbars}
\end{figure}

{\it Results and Discussion:}  We present our work in two parts (i) Determination of $f(z_i)$ from $f_8(z)$ data and background quantities and (ii) the constraints on 21-cm bias at these redshifts from measurement of $\beta_T$ is a radio interferometric intensity mapping experiment.

Figure \ref{fig:feightrecon} shows the reconstruction of $f_8(z)$ from data. In order to identify the redshift at which $f_8(z)$ peaks, we have obtained the MCMC \cite{foreman2013emcee} reconstruction of  $f_8(z)$ in two different ways. 
Case I: Without assuming any model and underlying Physics, we have considered 70 data points from diverse observational sources \cite{Kazantzidis_Perivolaropoulos_fs8_2018, Savvas_Nesseris_2012_fs8_SN_BAO} and simply fitted them using a fit function of the form the form 
$ f_8(z) = ( A_0 + A_1 z + A_2 z^2 )(1 + B_1 z + B_2 z^2)^{-1}$. 
In this approach, we are not interested in the interpretation of the expansion coefficients as much as we are keen on locating the peak of the function $f_8$.

Case II: Here we reconstruct $f_8$ from a weakly model-dependent semi-cosmographic approach \cite{P_Chavan_2025_semiCosmography, P_chavan_2025_semiCosmography_xpf8}. Starting from a Pad\'e rational fraction expansion in the angular diameter distance $D_A^{\cal P}(z, H_0, \alpha, \beta, \gamma)$ \cite{Saini_2000}, we obtain $H^{\cal P}{(z,\alpha, \beta, \gamma) = \left[d/dz(D_A^{\cal P}(1+z))\right]^{-1}}$. The dynamical equations in Eq: \ref{eq:autonomous_bgrnd_evln}, are solved using a semi-cosmographic equation of state dependent on a set of cosmographic parameters $( \alpha, \beta, \gamma)$, along with $\Omega_{m0}$, $\sigma_{8,0}$ and $H_0$. The $f_8(z)$ obtained from solving the dynamical system is then fitted with SDSS IV data for $ (D_M/r_d, D_H/r_d, D_V/r_d ~\text{and}~ f_8)$ by assuming CMBR priors on the sound horizon $r_d$ at the drag epoch. 
In this approach, we used SDSS data since the measurements of the RSD and BAO observables are simultaneously available.
The reconstructed $f_8$ obtained from the two methods indicates a $2.5\sigma $ tension between these two fits, and the tension is higher at larger redshifts. 
However, if we focus on the logarithmic slope $F_1$, then Fig. \ref{Fig: zeros_F1_F2_vs_z} shows that the zero crossing of $F_1(z)$ for the two different reconstruction methods in $z=z_1$ agrees quite well (despite the fact that the data and method used to reconstruct are completely different), indicating that the turning redshift for $f_8$ is robust and is not very sensitive to the data and reconstruction. We find that the zeros of $F_1$
corresponding to the peak in $f_8$ occur for case I and case II at
$z_1 = 0.412^{+0.167}_{-0.039}$ and $z_1 = 0.420^{+0.055}_{-0.050}$, respectively.

It is important to note that in reconstructing $f_8$, its local behaviour around the peak is crucial. The magnitude of the reconstructed function does not matter, but its turning point does, in this method.
In Fig. \ref{Fig: zeros_F1_F2_vs_z} we also show the behaviour of $F_2$, which is related to the double derivative of $f_8(z)$. The large errors in the reconstruction of $f_8$ especially at large redshifts in the model independent fit (due lack of data at large redshifts) leads to a disagreement between the best-fit values of zero crossings of $F_2$ from the two reconstructions, though they are within the large $1\sigma$ error bar of the former.
The reconstructed values of $z_2$ for case I and case II are
$z_2 = 1.159^{+0.818}_{-0.198}$ and $z_2 = 1.782^{+0.161}_{-0.151}$, respectively. The analytic properties of $f_8$ allow us to reconstruct $z_1$ and $z_2$.
These are the two redshifts where $f(z)$ can now be determined from the background quantities. 
In this work, the background quantities $H(z)$, $\Omega_{m0}$ and $w(z)$( related to $H$ and $dH/dz$) are reconstructed from SDSS IV data.
This choice is deliberate due to he availability of the covariance data between the RSD and BAO observables. 

Figure \ref{fig:zeros_F1_F2_vs_xi} demonstrates the graphical method to find roots of the equations $F_1( \xi_1, R(z_1), w(z_1)) = 0 $ and $F_2( \xi_2 , R(z_2), R'(z_2) , w(z_2), w'(z_2)) = 0$. 
The zero crossings $\xi_1$  and $\xi_2$ of these functions  give us $f(z_1) = \xi_1^{-1}$ and $f(z_2) = \xi_2^{-1}$ respectively. 
The errors in the reconstruction of $z_1$ and $z_2$ from the slopes of $f_8(z)$  along with the errors in the background quantities contribute to the final error in the measured values of $f(z_{1,2})$. 

The reconstructed $f(z)$ at the two redshifts is shown in Fig. \ref{fig:errorbars}. At $z=z_1$ the value of $f$ obtained from, the two methods for  fitting $f_8$ are given by $f(z_1)=0.718^{+0.107}_{-0.099}$ and $f(z_1) = 0.690^{+0.074}_{-0.071}$ for case I and II, respectively. The predicted value from the two methods has $<1\sigma$ discrepancy. The predicted projection from two sets of $\Lambda$CDM parameters (SDSS and Planck fitted) also agrees within $1\sigma$. 

The reconstruction of $f(z)$ at $z_2$ has large $z$ errors, and the projection from the semi-cosmographic approach with SDSS agrees much better than the blind polynomial fit with the $\Lambda$CDM. The projections on $f(z_2)$ for case I and II are $f(z_2)=1.083$ with lower bound $-0.162$ and $f(z_2)=0.919^{+0.032}_{-0.029}$, respectively.
With better quality of data at high redshifts and implementation of more robust model-agnostic methods like Gaussian reconstruction \cite{Holsclaw_2011_GPR,Shafieloo_2012_GPR, Jesus_2024_GPR, Dinda2024_GP_cosmography,Velazquez_mukherjee_GPR_2024, Purba_Mukherjee-Anjan_sen_GPR_2024, Dinda_2025} to fit the data, the projections are likely to improve.

We have discussed earlier that in the absence of knowledge about the bias, $f_8$ is measured in galaxy surveys. We have now seen that $f$ can be measured at specific redshifts using the galaxy clustering data itself.
We now discuss the prospect of measuring the post-reionization 21-cm bias at these redshifts from an observation of the 21-cm power spectrum at these redshifts.

The two redshifts where we have constrained $f(z)$  correspond to the observing frequencies of $\frac{1420 {~\rm MHz}}{(1+z_{1,2})} = 1000 {~\rm MHz}$, and $510.42 {~\rm MHz}$. We investigate the redshift where the error bar in the measurement of $f$ is smaller. If we consider the spherically averaged 21-cm power spectrum $ P_T(k) = C_T^2(z)  \left ( 1 + \frac{3}{2} \beta_T + \frac{1}{5}\beta_T^2 \right )  P_m(k,z)$. 
where the overall normalization $C_T$ in degenerate with $\sigma_8$, $b_T^2$ and $D^2_+(z)$. 
We consider a radio-interferometric observation whereby the power spectrum can be estimated from a visibility-visibility correlation \cite{poreion1, bali, Sarkar_2017, Bull_2015}.
By suitably averaging over visibility bins such that the covariance matrix is diagonal, the noise in the spherically averaged power spectrum is  given by  \cite{mcquinn2006cosmological} 
\begin{eqnarray} 
\delta P_T(k) &=& \left(  \sum _{\mu} \frac{1}{\delta P_T (k, \mu)^2} \right)^{-1/2} ~~~{\rm where} \nonumber \\
\delta P_{T}(k, \mu) &=& \frac{P_{T} + N_T}{\sqrt{N_c}} 
~{\text {with}} ~ N_T = \frac{\lambda^2 T_{sys}^2 r^2 {dr}/{d\nu}  }{A_e t_{\bf k}} 
\end{eqnarray} 
Here, for an observing redshift $z$, corresponding to a frequency $\nu$ and wavelength $\lambda = 21(1+z)$ cm, $r$ is the comoving distance to the source, $A_e$ is the effective area of the antenna dish, $B$ is the bandwidth and 
\[ N_c = 2 \pi k^2 \Delta k \Delta \mu ~ x^2  ({dx}/{d\nu})B \lambda^2 / A_e (2 \pi)^3 .\]
The system temperature $T_{sys}$ is assumed to be dominated by sky temperature. For a total observation time $T_0$, we have $t_{\bf k} = T_o N_{ant} ( N_{ant} -1 ) A_e \rho / 2 \lambda^2$, where $N_{ant}$ is the number of antennae in the array and $\rho$ is the normalized baseline distribution function.

We consider a SKA-1 Mid like radio interferometer with $197$ dish antennae, each of diameter $15$m and antenna efficiency of $0.7$. The array specifications are obtained from \footnote{https://www.skao.int/en}.
Considering $1000{~\rm hrs}$ observations and $10$ independent pointings at $\nu = 1000 {~\rm MHz}$ with a $128 {~\rm MHz}$ bandwidth, whereby the spherically averaged power spectrum can be measured at $\sim 10 \sigma$ at $k = 0.1 h ~\rm Mpc^{-1}$.
Considering $( C_T, \beta_T)$ as parameters of interest,  a Fisher matrix $$F_{mn} = \sum_k\frac{1} {\delta P_T(k)^2 }  \frac{\partial P_T}{\partial q_m}    \frac{\partial P_T}{\partial q_n}$$ gives errors on $\beta_T $ around its fiducial value from the Cramar-Rao bound. The bias is scale-dependent on small scales but is known to only depend on redshift on large scales. Assuming fiducial values of $(\beta_T, C_T)$ from the $\Lambda$CDM model \cite{PLANCK18_COSMO_params_Aghanim_2020} and bias model from simulations \cite{Sarkar_2016}, 
we find that the constraint on $f(z= 0.42)$ puts an error on the large scale $b_T(z=0.42, k < 0.01 ~\rm Mpc^{-1}) = 0.757 \pm 0.682$. 
This projected error on the large-scale bias assumes that foregrounds are completely removed. In principle, foregrounds shall severely degrade the projection \cite{di2002radio, Ghosh_2010, wang200621}, especially if many modes are lost due to the removal of the foreground wedge\cite{Liu-formalism1, liu2012well, liu2009improved}. 

We conclude by noting that by reconstructing the shape of the curve $f_8(z)$ and its derivatives, one can directly obtain $f(z)$ from purely background quantities. Thus, the RSD and BAO data can directly measure $f(z)$ at some specific redshifts. We have also demonstrated that this can constrain the large-scale 21-cm bias at some redshifts. This may have important consequences in the interpretation of the observed post-reionization 21-cm intensity maps in the future.

\bibliographystyle{apsrev4-2}
\bibliography{references}

\begin{thebibliography}{74}%
\makeatletter
\providecommand \@ifxundefined [1]{%
 \@ifx{#1\undefined}
}%
\providecommand \@ifnum [1]{%
 \ifnum #1\expandafter \@firstoftwo
 \else \expandafter \@secondoftwo
 \fi
}%
\providecommand \@ifx [1]{%
 \ifx #1\expandafter \@firstoftwo
 \else \expandafter \@secondoftwo
 \fi
}%
\providecommand \natexlab [1]{#1}%
\providecommand \enquote  [1]{``#1''}%
\providecommand \bibnamefont  [1]{#1}%
\providecommand \bibfnamefont [1]{#1}%
\providecommand \citenamefont [1]{#1}%
\providecommand \href@noop [0]{\@secondoftwo}%
\providecommand \href [0]{\begingroup \@sanitize@url \@href}%
\providecommand \@href[1]{\@@startlink{#1}\@@href}%
\providecommand \@@href[1]{\endgroup#1\@@endlink}%
\providecommand \@sanitize@url [0]{\catcode `\\12\catcode `\$12\catcode
  `\&12\catcode `\#12\catcode `\^12\catcode `\_12\catcode `\%12\relax}%
\providecommand \@@startlink[1]{}%
\providecommand \@@endlink[0]{}%
\providecommand \url  [0]{\begingroup\@sanitize@url \@url }%
\providecommand \@url [1]{\endgroup\@href {#1}{\urlprefix }}%
\providecommand \urlprefix  [0]{URL }%
\providecommand \Eprint [0]{\href }%
\providecommand \doibase [0]{https://doi.org/}%
\providecommand \selectlanguage [0]{\@gobble}%
\providecommand \bibinfo  [0]{\@secondoftwo}%
\providecommand \bibfield  [0]{\@secondoftwo}%
\providecommand \translation [1]{[#1]}%
\providecommand \BibitemOpen [0]{}%
\providecommand \bibitemStop [0]{}%
\providecommand \bibitemNoStop [0]{.\EOS\space}%
\providecommand \EOS [0]{\spacefactor3000\relax}%
\providecommand \BibitemShut  [1]{\csname bibitem#1\endcsname}%
\let\auto@bib@innerbib\@empty
%</preamble>
\bibitem [{\citenamefont {Sachs}\ and\ \citenamefont
  {Wolfe}(1986)}]{sachs1986perturbations}%
  \BibitemOpen
  \bibfield  {author} {\bibinfo {author} {\bibfnamefont {R.}~\bibnamefont
  {Sachs}}\ and\ \bibinfo {author} {\bibfnamefont {A.}~\bibnamefont {Wolfe}},\
  }\href@noop {} {\bibfield  {journal} {\bibinfo  {journal} {Inflationary
  Cosmology}\ ,\ \bibinfo {pages} {377}} (\bibinfo {year} {1986})}\BibitemShut
  {NoStop}%
\bibitem [{\citenamefont {{Bond}}\ \emph {et~al.}(1996)\citenamefont {{Bond}},
  \citenamefont {{Kofman}},\ and\ \citenamefont {{Pogosyan}}}]{bond-cosmicweb}%
  \BibitemOpen
  \bibfield  {author} {\bibinfo {author} {\bibfnamefont {J.~R.}\ \bibnamefont
  {{Bond}}}, \bibinfo {author} {\bibfnamefont {L.}~\bibnamefont {{Kofman}}},\
  and\ \bibinfo {author} {\bibfnamefont {D.}~\bibnamefont {{Pogosyan}}},\
  }\href {https://doi.org/10.1038/380603a0} {\bibfield  {journal} {\bibinfo
  {journal} {\nat}\ }\textbf {\bibinfo {volume} {380}},\ \bibinfo {pages} {603}
  (\bibinfo {year} {1996})},\ \Eprint {https://arxiv.org/abs/astro-ph/9512141}
  {arXiv:astro-ph/9512141 [astro-ph]} \BibitemShut {NoStop}%
\bibitem [{\citenamefont {{White}}\ \emph {et~al.}(1987)\citenamefont
  {{White}}, \citenamefont {{Frenk}}, \citenamefont {{Davis}},\ and\
  \citenamefont {{Efstathiou}}}]{White-cosmicweb}%
  \BibitemOpen
  \bibfield  {author} {\bibinfo {author} {\bibfnamefont {S.~D.~M.}\
  \bibnamefont {{White}}}, \bibinfo {author} {\bibfnamefont {C.~S.}\
  \bibnamefont {{Frenk}}}, \bibinfo {author} {\bibfnamefont {M.}~\bibnamefont
  {{Davis}}},\ and\ \bibinfo {author} {\bibfnamefont {G.}~\bibnamefont
  {{Efstathiou}}},\ }\href {https://doi.org/10.1086/164990} {\bibfield
  {journal} {\bibinfo  {journal} {\apj}\ }\textbf {\bibinfo {volume} {313}},\
  \bibinfo {pages} {505} (\bibinfo {year} {1987})}\BibitemShut {NoStop}%
\bibitem [{\citenamefont {Batista}(2021)}]{Batista_clustering-de}%
  \BibitemOpen
  \bibfield  {author} {\bibinfo {author} {\bibfnamefont {R.~C.}\ \bibnamefont
  {Batista}},\ }\href {https://doi.org/10.3390/universe8010022} {\bibfield
  {journal} {\bibinfo  {journal} {Universe}\ }\textbf {\bibinfo {volume} {8}},\
  \bibinfo {pages} {22} (\bibinfo {year} {2021})}\BibitemShut {NoStop}%
\bibitem [{\citenamefont {Maeder}(2017)}]{Maeder_2017_LCDM_alternative}%
  \BibitemOpen
  \bibfield  {author} {\bibinfo {author} {\bibfnamefont {A.}~\bibnamefont
  {Maeder}},\ }\href {https://doi.org/10.3847/1538-4357/834/2/194} {\bibfield
  {journal} {\bibinfo  {journal} {The Astrophysical Journal}\ }\textbf
  {\bibinfo {volume} {834}},\ \bibinfo {pages} {194} (\bibinfo {year}
  {2017})}\BibitemShut {NoStop}%
\bibitem [{\citenamefont {Bull}\ \emph {et~al.}(2016)\citenamefont {Bull},
  \citenamefont {Akrami}, \citenamefont {Adamek},\ and\ \citenamefont
  {et~al.}}]{BULL_et_al_2016_Beyond_LCDM}%
  \BibitemOpen
  \bibfield  {author} {\bibinfo {author} {\bibfnamefont {P.}~\bibnamefont
  {Bull}}, \bibinfo {author} {\bibfnamefont {Y.}~\bibnamefont {Akrami}},
  \bibinfo {author} {\bibfnamefont {J.}~\bibnamefont {Adamek}},\ and\ \bibinfo
  {author} {\bibnamefont {et~al.}},\ }\href
  {https://doi.org/https://doi.org/10.1016/j.dark.2016.02.001} {\bibfield
  {journal} {\bibinfo  {journal} {Physics of the Dark Universe}\ }\textbf
  {\bibinfo {volume} {12}},\ \bibinfo {pages} {56} (\bibinfo {year}
  {2016})}\BibitemShut {NoStop}%
\bibitem [{\citenamefont {Arun}\ \emph {et~al.}(2017)\citenamefont {Arun},
  \citenamefont {Gudennavar},\ and\ \citenamefont
  {Sivaram}}]{Arun_2017_DM_DE_alternative_models}%
  \BibitemOpen
  \bibfield  {author} {\bibinfo {author} {\bibfnamefont {K.}~\bibnamefont
  {Arun}}, \bibinfo {author} {\bibfnamefont {S.}~\bibnamefont {Gudennavar}},\
  and\ \bibinfo {author} {\bibfnamefont {C.}~\bibnamefont {Sivaram}},\ }\href
  {https://doi.org/https://doi.org/10.1016/j.asr.2017.03.043} {\bibfield
  {journal} {\bibinfo  {journal} {Advances in Space Research}\ }\textbf
  {\bibinfo {volume} {60}},\ \bibinfo {pages} {166} (\bibinfo {year}
  {2017})}\BibitemShut {NoStop}%
\bibitem [{\citenamefont {Hou}\ \emph {et~al.}(2020)\citenamefont {Hou},
  \citenamefont {Sánchez}, \citenamefont {Ross}, \citenamefont {Smith},\ and\
  \citenamefont {Neveux}}]{Hou_2020}%
  \BibitemOpen
  \bibfield  {author} {\bibinfo {author} {\bibfnamefont {J.}~\bibnamefont
  {Hou}}, \bibinfo {author} {\bibfnamefont {A.~G.}\ \bibnamefont {Sánchez}},
  \bibinfo {author} {\bibfnamefont {A.~J.}\ \bibnamefont {Ross}}, \bibinfo
  {author} {\bibfnamefont {A.}~\bibnamefont {Smith}},\ and\ \bibinfo {author}
  {\bibfnamefont {e.}~\bibnamefont {Neveux}},\ }\href
  {https://doi.org/10.1093/mnras/staa3234} {\bibfield  {journal} {\bibinfo
  {journal} {Monthly Notices of the Royal Astronomical Society}\ }\textbf
  {\bibinfo {volume} {500}},\ \bibinfo {pages} {1201–1221} (\bibinfo {year}
  {2020})}\BibitemShut {NoStop}%
\bibitem [{\citenamefont {Beutler}\ \emph {et~al.}(2012)\citenamefont
  {Beutler}, \citenamefont {Blake}, \citenamefont {Colless}, \citenamefont
  {Jones}, \citenamefont {Staveley-Smith}, \citenamefont {Poole}, \citenamefont
  {Campbell}, \citenamefont {Parker}, \citenamefont {Saunders},\ and\
  \citenamefont {Watson}}]{Beutler_2012_6dF_growthRate}%
  \BibitemOpen
  \bibfield  {author} {\bibinfo {author} {\bibfnamefont {F.}~\bibnamefont
  {Beutler}}, \bibinfo {author} {\bibfnamefont {C.}~\bibnamefont {Blake}},
  \bibinfo {author} {\bibfnamefont {M.}~\bibnamefont {Colless}}, \bibinfo
  {author} {\bibfnamefont {D.~H.}\ \bibnamefont {Jones}}, \bibinfo {author}
  {\bibfnamefont {L.}~\bibnamefont {Staveley-Smith}}, \bibinfo {author}
  {\bibfnamefont {G.~B.}\ \bibnamefont {Poole}}, \bibinfo {author}
  {\bibfnamefont {L.}~\bibnamefont {Campbell}}, \bibinfo {author}
  {\bibfnamefont {Q.}~\bibnamefont {Parker}}, \bibinfo {author} {\bibfnamefont
  {W.}~\bibnamefont {Saunders}},\ and\ \bibinfo {author} {\bibfnamefont
  {F.}~\bibnamefont {Watson}},\ }\href
  {https://doi.org/10.1111/j.1365-2966.2012.21136.x} {\bibfield  {journal}
  {\bibinfo  {journal} {Monthly Notices of the Royal Astronomical Society}\
  }\textbf {\bibinfo {volume} {423}},\ \bibinfo {pages} {3430} (\bibinfo {year}
  {2012})},\ \Eprint
  {https://arxiv.org/abs/https://academic.oup.com/mnras/article-pdf/423/4/3430/4903419/mnras0423-3430.pdf}
  {https://academic.oup.com/mnras/article-pdf/423/4/3430/4903419/mnras0423-3430.pdf}
  \BibitemShut {NoStop}%
\bibitem [{\citenamefont {Alam}\ and\ \citenamefont
  {et~al.}(2017)}]{Alam_2017_SDSS_III}%
  \BibitemOpen
  \bibfield  {author} {\bibinfo {author} {\bibfnamefont {S.}~\bibnamefont
  {Alam}}\ and\ \bibinfo {author} {\bibnamefont {et~al.}},\ }\href
  {https://doi.org/10.1093/mnras/stx721} {\bibfield  {journal} {\bibinfo
  {journal} {Monthly Notices of the Royal Astronomical Society}\ }\textbf
  {\bibinfo {volume} {470}},\ \bibinfo {pages} {2617} (\bibinfo {year}
  {2017})},\ \Eprint
  {https://arxiv.org/abs/https://academic.oup.com/mnras/article-pdf/470/3/2617/18315003/stx721.pdf}
  {https://academic.oup.com/mnras/article-pdf/470/3/2617/18315003/stx721.pdf}
  \BibitemShut {NoStop}%
\bibitem [{\citenamefont {Howlett}\ \emph {et~al.}(2015)\citenamefont
  {Howlett}, \citenamefont {Ross}, \citenamefont {Samushia}, \citenamefont
  {Percival},\ and\ \citenamefont {Manera}}]{Howlett_2015_fs8}%
  \BibitemOpen
  \bibfield  {author} {\bibinfo {author} {\bibfnamefont {C.}~\bibnamefont
  {Howlett}}, \bibinfo {author} {\bibfnamefont {A.~J.}\ \bibnamefont {Ross}},
  \bibinfo {author} {\bibfnamefont {L.}~\bibnamefont {Samushia}}, \bibinfo
  {author} {\bibfnamefont {W.~J.}\ \bibnamefont {Percival}},\ and\ \bibinfo
  {author} {\bibfnamefont {M.}~\bibnamefont {Manera}},\ }\href
  {https://doi.org/10.1093/mnras/stu2693} {\bibfield  {journal} {\bibinfo
  {journal} {Monthly Notices of the Royal Astronomical Society}\ }\textbf
  {\bibinfo {volume} {449}},\ \bibinfo {pages} {848} (\bibinfo {year}
  {2015})},\ \Eprint
  {https://arxiv.org/abs/https://academic.oup.com/mnras/article-pdf/449/1/848/17335801/stu2693.pdf}
  {https://academic.oup.com/mnras/article-pdf/449/1/848/17335801/stu2693.pdf}
  \BibitemShut {NoStop}%
\bibitem [{\citenamefont {Percival}\ \emph {et~al.}(2007)\citenamefont
  {Percival}, \citenamefont {Cole}, \citenamefont {Eisenstein}, \citenamefont
  {Nichol}, \citenamefont {Peacock}, \citenamefont {Pope},\ and\ \citenamefont
  {Szalay}}]{Percival_2007}%
  \BibitemOpen
  \bibfield  {author} {\bibinfo {author} {\bibfnamefont {W.~J.}\ \bibnamefont
  {Percival}}, \bibinfo {author} {\bibfnamefont {S.}~\bibnamefont {Cole}},
  \bibinfo {author} {\bibfnamefont {D.~J.}\ \bibnamefont {Eisenstein}},
  \bibinfo {author} {\bibfnamefont {R.~C.}\ \bibnamefont {Nichol}}, \bibinfo
  {author} {\bibfnamefont {J.~A.}\ \bibnamefont {Peacock}}, \bibinfo {author}
  {\bibfnamefont {A.~C.}\ \bibnamefont {Pope}},\ and\ \bibinfo {author}
  {\bibfnamefont {A.~S.}\ \bibnamefont {Szalay}},\ }\href
  {https://doi.org/10.1111/j.1365-2966.2007.12268.x} {\bibfield  {journal}
  {\bibinfo  {journal} {Monthly Notices of the Royal Astronomical Society}\
  }\textbf {\bibinfo {volume} {381}},\ \bibinfo {pages} {1053–1066} (\bibinfo
  {year} {2007})}\BibitemShut {NoStop}%
\bibitem [{\citenamefont {McDonald}\ \emph {et~al.}(2006)\citenamefont
  {McDonald}, \citenamefont {Seljak}, \citenamefont {Burles}, \citenamefont
  {Schlegel}, \citenamefont {Weinberg}, \citenamefont {Cen}, \citenamefont
  {Shih}, \citenamefont {Schaye}, \citenamefont {Schneider}, \citenamefont
  {Bahcall} \emph {et~al.}}]{mcdonald2006lyalpha}%
  \BibitemOpen
  \bibfield  {author} {\bibinfo {author} {\bibfnamefont {P.}~\bibnamefont
  {McDonald}}, \bibinfo {author} {\bibfnamefont {U.}~\bibnamefont {Seljak}},
  \bibinfo {author} {\bibfnamefont {S.}~\bibnamefont {Burles}}, \bibinfo
  {author} {\bibfnamefont {D.~J.}\ \bibnamefont {Schlegel}}, \bibinfo {author}
  {\bibfnamefont {D.~H.}\ \bibnamefont {Weinberg}}, \bibinfo {author}
  {\bibfnamefont {R.}~\bibnamefont {Cen}}, \bibinfo {author} {\bibfnamefont
  {D.}~\bibnamefont {Shih}}, \bibinfo {author} {\bibfnamefont {J.}~\bibnamefont
  {Schaye}}, \bibinfo {author} {\bibfnamefont {D.~P.}\ \bibnamefont
  {Schneider}}, \bibinfo {author} {\bibfnamefont {N.~A.}\ \bibnamefont
  {Bahcall}}, \emph {et~al.},\ }\href {https://doi.org/10.1086/444361}
  {\bibfield  {journal} {\bibinfo  {journal} {The Astrophysical Journal
  Supplement Series}\ }\textbf {\bibinfo {volume} {163}},\ \bibinfo {pages}
  {80} (\bibinfo {year} {2006})}\BibitemShut {NoStop}%
\bibitem [{\citenamefont {Ir{\v{s}}i{\v{c}}}\ \emph {et~al.}(2017)\citenamefont
  {Ir{\v{s}}i{\v{c}}}, \citenamefont {Viel}, \citenamefont {Berg},
  \citenamefont {D'Odorico}, \citenamefont {Haehnelt}, \citenamefont
  {Cristiani}, \citenamefont {Cupani}, \citenamefont {Kim}, \citenamefont
  {L{\'o}pez}, \citenamefont {Ellison} \emph {et~al.}}]{irvsivc2017lyman}%
  \BibitemOpen
  \bibfield  {author} {\bibinfo {author} {\bibfnamefont {V.}~\bibnamefont
  {Ir{\v{s}}i{\v{c}}}}, \bibinfo {author} {\bibfnamefont {M.}~\bibnamefont
  {Viel}}, \bibinfo {author} {\bibfnamefont {T.~A.}\ \bibnamefont {Berg}},
  \bibinfo {author} {\bibfnamefont {V.}~\bibnamefont {D'Odorico}}, \bibinfo
  {author} {\bibfnamefont {M.~G.}\ \bibnamefont {Haehnelt}}, \bibinfo {author}
  {\bibfnamefont {S.}~\bibnamefont {Cristiani}}, \bibinfo {author}
  {\bibfnamefont {G.}~\bibnamefont {Cupani}}, \bibinfo {author} {\bibfnamefont
  {T.-S.}\ \bibnamefont {Kim}}, \bibinfo {author} {\bibfnamefont
  {S.}~\bibnamefont {L{\'o}pez}}, \bibinfo {author} {\bibfnamefont
  {S.}~\bibnamefont {Ellison}}, \emph {et~al.},\ }\href
  {https://doi.org/10.1093/mnras/stw3372} {\bibfield  {journal} {\bibinfo
  {journal} {Monthly Notices of the Royal Astronomical Society}\ }\textbf
  {\bibinfo {volume} {466}},\ \bibinfo {pages} {4332} (\bibinfo {year}
  {2017})}\BibitemShut {NoStop}%
\bibitem [{\citenamefont {Slosar}\ \emph {et~al.}(2011)\citenamefont {Slosar},
  \citenamefont {Font-Ribera}, \citenamefont {Pieri}, \citenamefont {Rich},
  \citenamefont {Goff}, \citenamefont {Aubourg}, \citenamefont {Brinkmann},
  \citenamefont {Busca}, \citenamefont {Carithers}, \citenamefont
  {Charlassier},\ and\ \citenamefont {et~al.}}]{Slosar_2011}%
  \BibitemOpen
  \bibfield  {author} {\bibinfo {author} {\bibfnamefont {A.}~\bibnamefont
  {Slosar}}, \bibinfo {author} {\bibfnamefont {A.}~\bibnamefont {Font-Ribera}},
  \bibinfo {author} {\bibfnamefont {M.~M.}\ \bibnamefont {Pieri}}, \bibinfo
  {author} {\bibfnamefont {J.}~\bibnamefont {Rich}}, \bibinfo {author}
  {\bibfnamefont {J.-M.~L.}\ \bibnamefont {Goff}}, \bibinfo {author}
  {\bibfnamefont {E.}~\bibnamefont {Aubourg}}, \bibinfo {author} {\bibfnamefont
  {J.}~\bibnamefont {Brinkmann}}, \bibinfo {author} {\bibfnamefont
  {N.}~\bibnamefont {Busca}}, \bibinfo {author} {\bibfnamefont
  {B.}~\bibnamefont {Carithers}}, \bibinfo {author} {\bibfnamefont
  {R.}~\bibnamefont {Charlassier}},\ and\ \bibinfo {author} {\bibnamefont
  {et~al.}},\ }\href {https://doi.org/10.1088/1475-7516/2011/09/001} {\bibfield
   {journal} {\bibinfo  {journal} {Journal of Cosmology and Astroparticle
  Physics}\ }\textbf {\bibinfo {volume} {2011}}\bibinfo  {number} { (09)},\
  \bibinfo {pages} {001–001}}\BibitemShut {NoStop}%
\bibitem [{\citenamefont {Garzilli}\ \emph {et~al.}(2019)\citenamefont
  {Garzilli}, \citenamefont {Magalich}, \citenamefont {Theuns}, \citenamefont
  {Frenk}, \citenamefont {Weniger}, \citenamefont {Ruchayskiy},\ and\
  \citenamefont {Boyarsky}}]{Garzilli_2019}%
  \BibitemOpen
\bibfield  {number} {  }\bibfield  {author} {\bibinfo {author} {\bibfnamefont
  {A.}~\bibnamefont {Garzilli}}, \bibinfo {author} {\bibfnamefont
  {A.}~\bibnamefont {Magalich}}, \bibinfo {author} {\bibfnamefont
  {T.}~\bibnamefont {Theuns}}, \bibinfo {author} {\bibfnamefont {C.~S.}\
  \bibnamefont {Frenk}}, \bibinfo {author} {\bibfnamefont {C.}~\bibnamefont
  {Weniger}}, \bibinfo {author} {\bibfnamefont {O.}~\bibnamefont
  {Ruchayskiy}},\ and\ \bibinfo {author} {\bibfnamefont {A.}~\bibnamefont
  {Boyarsky}},\ }\href {https://doi.org/10.1093/mnras/stz2188} {\bibfield
  {journal} {\bibinfo  {journal} {Monthly Notices of the Royal Astronomical
  Society}\ }\textbf {\bibinfo {volume} {489}},\ \bibinfo {pages} {3456–3471}
  (\bibinfo {year} {2019})}\BibitemShut {NoStop}%
\bibitem [{\citenamefont {Hamilton}(1998)}]{hamilton1998linear}%
  \BibitemOpen
  \bibfield  {author} {\bibinfo {author} {\bibfnamefont {A.}~\bibnamefont
  {Hamilton}},\ }in\ \href {https://doi.org/10.1007/978-94-011-4960-0_17}
  {\emph {\bibinfo {booktitle} {The evolving universe}}}\ (\bibinfo
  {publisher} {Springer},\ \bibinfo {year} {1998})\ pp.\ \bibinfo {pages}
  {185--275}\BibitemShut {NoStop}%
\bibitem [{\citenamefont {Delubac}\ \emph {et~al.}(2015)\citenamefont
  {Delubac}, \citenamefont {Bautista}, \citenamefont {Busca}, \citenamefont
  {Rich}, \citenamefont {Kirkby}, \citenamefont {Bailey}, \citenamefont
  {Font-Ribera}, \citenamefont {Slosar}, \citenamefont {Lee}, \citenamefont
  {Pieri},\ and\ \citenamefont {et~al.}}]{Delubac_2015}%
  \BibitemOpen
  \bibfield  {author} {\bibinfo {author} {\bibfnamefont {T.}~\bibnamefont
  {Delubac}}, \bibinfo {author} {\bibfnamefont {J.~E.}\ \bibnamefont
  {Bautista}}, \bibinfo {author} {\bibfnamefont {N.~G.}\ \bibnamefont {Busca}},
  \bibinfo {author} {\bibfnamefont {J.}~\bibnamefont {Rich}}, \bibinfo {author}
  {\bibfnamefont {D.}~\bibnamefont {Kirkby}}, \bibinfo {author} {\bibfnamefont
  {S.}~\bibnamefont {Bailey}}, \bibinfo {author} {\bibfnamefont
  {A.}~\bibnamefont {Font-Ribera}}, \bibinfo {author} {\bibfnamefont
  {A.}~\bibnamefont {Slosar}}, \bibinfo {author} {\bibfnamefont {K.-G.}\
  \bibnamefont {Lee}}, \bibinfo {author} {\bibfnamefont {M.~M.}\ \bibnamefont
  {Pieri}},\ and\ \bibinfo {author} {\bibnamefont {et~al.}},\ }\bibfield
  {journal} {\bibinfo  {journal} {Astronomy and Astrophysics}\ }\textbf
  {\bibinfo {volume} {574}},\ \href
  {https://doi.org/10.1051/0004-6361/201423969} {10.1051/0004-6361/201423969}
  (\bibinfo {year} {2015})\BibitemShut {NoStop}%
\bibitem [{\citenamefont {Font-Ribera}\ \emph {et~al.}(2012)\citenamefont
  {Font-Ribera}, \citenamefont {Miralda-Escudé}, \citenamefont {Arnau},
  \citenamefont {Carithers}, \citenamefont {Lee}, \citenamefont {Noterdaeme},
  \citenamefont {Pâris}, \citenamefont {Petitjean}, \citenamefont {Rich},
  \citenamefont {Rollinde},\ and\ \citenamefont {et~al.}}]{FontRibera_2012}%
  \BibitemOpen
  \bibfield  {author} {\bibinfo {author} {\bibfnamefont {A.}~\bibnamefont
  {Font-Ribera}}, \bibinfo {author} {\bibfnamefont {J.}~\bibnamefont
  {Miralda-Escudé}}, \bibinfo {author} {\bibfnamefont {E.}~\bibnamefont
  {Arnau}}, \bibinfo {author} {\bibfnamefont {B.}~\bibnamefont {Carithers}},
  \bibinfo {author} {\bibfnamefont {K.-G.}\ \bibnamefont {Lee}}, \bibinfo
  {author} {\bibfnamefont {P.}~\bibnamefont {Noterdaeme}}, \bibinfo {author}
  {\bibfnamefont {I.}~\bibnamefont {Pâris}}, \bibinfo {author} {\bibfnamefont
  {P.}~\bibnamefont {Petitjean}}, \bibinfo {author} {\bibfnamefont
  {J.}~\bibnamefont {Rich}}, \bibinfo {author} {\bibfnamefont {E.}~\bibnamefont
  {Rollinde}},\ and\ \bibinfo {author} {\bibnamefont {et~al.}},\ }\href
  {https://doi.org/10.1088/1475-7516/2012/11/059} {\bibfield  {journal}
  {\bibinfo  {journal} {Journal of Cosmology and Astroparticle Physics}\
  }\textbf {\bibinfo {volume} {2012}}\bibinfo  {number} { (11)}}\BibitemShut
  {NoStop}%
\bibitem [{\citenamefont {Zhao}\ \emph {et~al.}(2016)\citenamefont {Zhao},
  \citenamefont {Wang}, \citenamefont {Ross}, \citenamefont {Shandera},
  \citenamefont {Percival}, \citenamefont {Dawson}, \citenamefont {Kneib},
  \citenamefont {Myers}, \citenamefont {Brownstein}, \citenamefont {Comparat},\
  and\ \citenamefont {et~al.}}]{Zhao_2016_eBOSS}%
  \BibitemOpen
\bibfield  {number} {  }\bibfield  {author} {\bibinfo {author} {\bibfnamefont
  {G.-B.}\ \bibnamefont {Zhao}}, \bibinfo {author} {\bibfnamefont
  {Y.}~\bibnamefont {Wang}}, \bibinfo {author} {\bibfnamefont {A.~J.}\
  \bibnamefont {Ross}}, \bibinfo {author} {\bibfnamefont {S.}~\bibnamefont
  {Shandera}}, \bibinfo {author} {\bibfnamefont {W.~J.}\ \bibnamefont
  {Percival}}, \bibinfo {author} {\bibfnamefont {K.~S.}\ \bibnamefont
  {Dawson}}, \bibinfo {author} {\bibfnamefont {J.-P.}\ \bibnamefont {Kneib}},
  \bibinfo {author} {\bibfnamefont {A.~D.}\ \bibnamefont {Myers}}, \bibinfo
  {author} {\bibfnamefont {J.~R.}\ \bibnamefont {Brownstein}}, \bibinfo
  {author} {\bibfnamefont {J.}~\bibnamefont {Comparat}},\ and\ \bibinfo
  {author} {\bibnamefont {et~al.}},\ }\href
  {https://doi.org/10.1093/mnras/stw135} {\bibfield  {journal} {\bibinfo
  {journal} {Monthly Notices of the Royal Astronomical Society}\ }\textbf
  {\bibinfo {volume} {457}},\ \bibinfo {pages} {2377–2390} (\bibinfo {year}
  {2016})}\BibitemShut {NoStop}%
\bibitem [{\citenamefont {Ivanov}(2021)}]{Ivanov_2021_elg}%
  \BibitemOpen
  \bibfield  {author} {\bibinfo {author} {\bibfnamefont {M.~M.}\ \bibnamefont
  {Ivanov}},\ }\href {https://doi.org/10.1103/PhysRevD.104.103514} {\bibfield
  {journal} {\bibinfo  {journal} {Phys. Rev. D}\ }\textbf {\bibinfo {volume}
  {104}},\ \bibinfo {pages} {103514} (\bibinfo {year} {2021})}\BibitemShut
  {NoStop}%
\bibitem [{\citenamefont {Collaboration}(2025)}]{DESI_Colab_2025_DR2}%
  \BibitemOpen
  \bibfield  {author} {\bibinfo {author} {\bibfnamefont {D.}~\bibnamefont
  {Collaboration}},\ }\href {https://arxiv.org/abs/2503.14738} {\bibinfo
  {title} {Desi dr2 results ii: Measurements of baryon acoustic oscillations
  and cosmological constraints}} (\bibinfo {year} {2025}),\ \Eprint
  {https://arxiv.org/abs/2503.14738} {arXiv:2503.14738 [astro-ph.CO]}
  \BibitemShut {NoStop}%
\bibitem [{\citenamefont {Adame}\ \emph {et~al.}(2025)\citenamefont {Adame},
  \citenamefont {Aguilar}, \citenamefont {Ahlen}, \citenamefont {Alam},\ and\
  \citenamefont {collaboration}}]{Adame_DESI_Data_2025}%
  \BibitemOpen
  \bibfield  {author} {\bibinfo {author} {\bibfnamefont {A.}~\bibnamefont
  {Adame}}, \bibinfo {author} {\bibfnamefont {J.}~\bibnamefont {Aguilar}},
  \bibinfo {author} {\bibfnamefont {S.}~\bibnamefont {Ahlen}}, \bibinfo
  {author} {\bibfnamefont {S.~e.}\ \bibnamefont {Alam}},\ and\ \bibinfo
  {author} {\bibfnamefont {T.~D.}\ \bibnamefont {collaboration}},\ }\href
  {https://doi.org/10.1088/1475-7516/2025/02/021} {\bibfield  {journal}
  {\bibinfo  {journal} {Journal of Cosmology and Astroparticle Physics}\
  }\textbf {\bibinfo {volume} {2025}}\bibinfo  {number} { (02)},\ \bibinfo
  {pages} {021}}\BibitemShut {NoStop}%
\bibitem [{\citenamefont {Levi}\ \emph {et~al.}(2013)\citenamefont {Levi},
  \citenamefont {Bebek}, \citenamefont {Beers}, \citenamefont {Blum},
  \citenamefont {Cahn}, \citenamefont {Eisenstein}, \citenamefont {Flaugher},
  \citenamefont {Honscheid}, \citenamefont {Kron}, \citenamefont {Lahav} \emph
  {et~al.}}]{levi2013desi}%
  \BibitemOpen
\bibfield  {number} {  }\bibfield  {author} {\bibinfo {author} {\bibfnamefont
  {M.}~\bibnamefont {Levi}}, \bibinfo {author} {\bibfnamefont {C.}~\bibnamefont
  {Bebek}}, \bibinfo {author} {\bibfnamefont {T.}~\bibnamefont {Beers}},
  \bibinfo {author} {\bibfnamefont {R.}~\bibnamefont {Blum}}, \bibinfo {author}
  {\bibfnamefont {R.}~\bibnamefont {Cahn}}, \bibinfo {author} {\bibfnamefont
  {D.}~\bibnamefont {Eisenstein}}, \bibinfo {author} {\bibfnamefont
  {B.}~\bibnamefont {Flaugher}}, \bibinfo {author} {\bibfnamefont
  {K.}~\bibnamefont {Honscheid}}, \bibinfo {author} {\bibfnamefont
  {R.}~\bibnamefont {Kron}}, \bibinfo {author} {\bibfnamefont {O.}~\bibnamefont
  {Lahav}}, \emph {et~al.},\ }\href {https://arxiv.org/abs/1308.0847}
  {\bibfield  {journal} {\bibinfo  {journal} {arXiv preprint arXiv:1308.0847}\
  } (\bibinfo {year} {2013})},\ \Eprint {https://arxiv.org/abs/1308.0847}
  {arXiv:1308.0847 [astro-ph.CO]} \BibitemShut {NoStop}%
\bibitem [{\citenamefont {Scaramella}\ \emph {et~al.}(2022)\citenamefont
  {Scaramella}, \citenamefont {Amiaux}, \citenamefont {Mellier}, \citenamefont
  {Burigana}, \citenamefont {Carvalho}, \citenamefont {Cuillandre},
  \citenamefont {Da~Silva}, \citenamefont {Derosa}, \citenamefont {Dinis},
  \citenamefont {Maiorano} \emph {et~al.}}]{Scaramella_2022_euclid_prepI}%
  \BibitemOpen
  \bibfield  {author} {\bibinfo {author} {\bibfnamefont {R.}~\bibnamefont
  {Scaramella}}, \bibinfo {author} {\bibfnamefont {J.}~\bibnamefont {Amiaux}},
  \bibinfo {author} {\bibfnamefont {Y.}~\bibnamefont {Mellier}}, \bibinfo
  {author} {\bibfnamefont {C.}~\bibnamefont {Burigana}}, \bibinfo {author}
  {\bibfnamefont {C.}~\bibnamefont {Carvalho}}, \bibinfo {author}
  {\bibfnamefont {J.-C.}\ \bibnamefont {Cuillandre}}, \bibinfo {author}
  {\bibfnamefont {A.}~\bibnamefont {Da~Silva}}, \bibinfo {author}
  {\bibfnamefont {A.}~\bibnamefont {Derosa}}, \bibinfo {author} {\bibfnamefont
  {J.}~\bibnamefont {Dinis}}, \bibinfo {author} {\bibfnamefont
  {E.}~\bibnamefont {Maiorano}}, \emph {et~al.},\ }\href
  {https://doi.org/10.1051/0004-6361/202141938} {\bibfield  {journal} {\bibinfo
   {journal} {Astronomy \& Astrophysics}\ }\textbf {\bibinfo {volume} {662}},\
  \bibinfo {pages} {A112} (\bibinfo {year} {2022})}\BibitemShut {NoStop}%
\bibitem [{\citenamefont {{Planck
  Collaboration}}(2020)}]{PLANCK18_COSMO_params_Aghanim_2020}%
  \BibitemOpen
  \bibfield  {author} {\bibinfo {author} {\bibnamefont {{Planck
  Collaboration}}},\ }\href {https://doi.org/10.1051/0004-6361/201833910}
  {\bibfield  {journal} {\bibinfo  {journal} {A \& A}\ }\textbf {\bibinfo
  {volume} {641}},\ \bibinfo {pages} {A6} (\bibinfo {year} {2020})}\BibitemShut
  {NoStop}%
\bibitem [{\citenamefont {Abbott}\ \emph {et~al.}(2022)\citenamefont {Abbott},
  \citenamefont {Aguena}, \citenamefont {Alarcon}, \citenamefont {Allam},
  \citenamefont {Alves}, \citenamefont {Amon}, \citenamefont
  {Andrade-Oliveira}, \citenamefont {Annis}, \citenamefont {Avila},
  \citenamefont {Bacon} \emph {et~al.}}]{abbott2022_DESY_weaklensing}%
  \BibitemOpen
  \bibfield  {author} {\bibinfo {author} {\bibfnamefont {T.}~\bibnamefont
  {Abbott}}, \bibinfo {author} {\bibfnamefont {M.}~\bibnamefont {Aguena}},
  \bibinfo {author} {\bibfnamefont {A.}~\bibnamefont {Alarcon}}, \bibinfo
  {author} {\bibfnamefont {S.}~\bibnamefont {Allam}}, \bibinfo {author}
  {\bibfnamefont {O.}~\bibnamefont {Alves}}, \bibinfo {author} {\bibfnamefont
  {A.}~\bibnamefont {Amon}}, \bibinfo {author} {\bibfnamefont {F.}~\bibnamefont
  {Andrade-Oliveira}}, \bibinfo {author} {\bibfnamefont {J.}~\bibnamefont
  {Annis}}, \bibinfo {author} {\bibfnamefont {S.}~\bibnamefont {Avila}},
  \bibinfo {author} {\bibfnamefont {D.}~\bibnamefont {Bacon}}, \emph {et~al.},\
  }\href {https://doi.org/10.1103/PhysRevD.105.023520} {\bibfield  {journal}
  {\bibinfo  {journal} {Physical Review D}\ }\textbf {\bibinfo {volume}
  {105}},\ \bibinfo {pages} {023520} (\bibinfo {year} {2022})}\BibitemShut
  {NoStop}%
\bibitem [{\citenamefont {Porredon}\ and\ \citenamefont
  {et~al.}(2022)}]{Porredon_DESY_lensing_2022}%
  \BibitemOpen
  \bibfield  {author} {\bibinfo {author} {\bibfnamefont {A.}~\bibnamefont
  {Porredon}}\ and\ \bibinfo {author} {\bibnamefont {et~al.}} (\bibinfo
  {collaboration} {DES Collaboration}),\ }\href
  {https://doi.org/10.1103/PhysRevD.106.103530} {\bibfield  {journal} {\bibinfo
   {journal} {Phys. Rev. D}\ }\textbf {\bibinfo {volume} {106}},\ \bibinfo
  {pages} {103530} (\bibinfo {year} {2022})}\BibitemShut {NoStop}%
\bibitem [{\citenamefont {Pandey}\ and\ \citenamefont
  {et~al.}(2022)}]{Pandey_DESY_gal_gal_lensing_2022}%
  \BibitemOpen
  \bibfield  {author} {\bibinfo {author} {\bibfnamefont {S.}~\bibnamefont
  {Pandey}}\ and\ \bibinfo {author} {\bibnamefont {et~al.}} (\bibinfo
  {collaboration} {DES Collaboration}),\ }\href
  {https://doi.org/10.1103/PhysRevD.106.043520} {\bibfield  {journal} {\bibinfo
   {journal} {Phys. Rev. D}\ }\textbf {\bibinfo {volume} {106}},\ \bibinfo
  {pages} {043520} (\bibinfo {year} {2022})}\BibitemShut {NoStop}%
\bibitem [{\citenamefont {Hikage}\ and\ \citenamefont
  {et~al.}(2019)}]{Hikage_2019_CosmicShear_PS}%
  \BibitemOpen
  \bibfield  {author} {\bibinfo {author} {\bibfnamefont {C.}~\bibnamefont
  {Hikage}}\ and\ \bibinfo {author} {\bibnamefont {et~al.}},\ }\href
  {https://doi.org/10.1093/pasj/psz010} {\bibfield  {journal} {\bibinfo
  {journal} {Publications of the Astronomical Society of Japan}\ }\textbf
  {\bibinfo {volume} {71}},\ \bibinfo {pages} {43} (\bibinfo {year} {2019})},\
  \Eprint
  {https://arxiv.org/abs/https://academic.oup.com/pasj/article-pdf/71/2/43/54666032/pasj\_71\_2\_43.pdf}
  {https://academic.oup.com/pasj/article-pdf/71/2/43/54666032/pasj\_71\_2\_43.pdf}
  \BibitemShut {NoStop}%
\bibitem [{\citenamefont {Karim}\ and\ \citenamefont
  {et~al.}(2025)}]{Karim_2025_sigma8}%
  \BibitemOpen
  \bibfield  {author} {\bibinfo {author} {\bibfnamefont {T.}~\bibnamefont
  {Karim}}\ and\ \bibinfo {author} {\bibnamefont {et~al.}},\ }\href
  {https://doi.org/10.1088/1475-7516/2025/02/045} {\bibfield  {journal}
  {\bibinfo  {journal} {Journal of Cosmology and Astroparticle Physics}\
  }\textbf {\bibinfo {volume} {2025}}\bibinfo  {number} { (02)},\ \bibinfo
  {pages} {045}}\BibitemShut {NoStop}%
\bibitem [{\citenamefont {Nusser}\ \emph {et~al.}(2011)\citenamefont {Nusser},
  \citenamefont {Branchini},\ and\ \citenamefont
  {Davis}}]{Nusser_2012_log_grth_rt}%
  \BibitemOpen
\bibfield  {number} {  }\bibfield  {author} {\bibinfo {author} {\bibfnamefont
  {A.}~\bibnamefont {Nusser}}, \bibinfo {author} {\bibfnamefont
  {E.}~\bibnamefont {Branchini}},\ and\ \bibinfo {author} {\bibfnamefont
  {M.}~\bibnamefont {Davis}},\ }\href
  {https://doi.org/10.1088/0004-637X/744/2/193} {\bibfield  {journal} {\bibinfo
   {journal} {The Astrophysical Journal}\ }\textbf {\bibinfo {volume} {744}},\
  \bibinfo {pages} {193} (\bibinfo {year} {2011})}\BibitemShut {NoStop}%
\bibitem [{\citenamefont {{Wyithe}}\ and\ \citenamefont
  {{Loeb}}(2009)}]{poreion0}%
  \BibitemOpen
  \bibfield  {author} {\bibinfo {author} {\bibfnamefont {J.~S.~B.}\
  \bibnamefont {{Wyithe}}}\ and\ \bibinfo {author} {\bibfnamefont
  {A.}~\bibnamefont {{Loeb}}},\ }\href
  {https://doi.org/10.1111/j.1365-2966.2009.15019.x} {\bibfield  {journal}
  {\bibinfo  {journal} {MNRAS}\ }\textbf {\bibinfo {volume} {397}},\ \bibinfo
  {pages} {1926} (\bibinfo {year} {2009})}\BibitemShut {NoStop}%
\bibitem [{\citenamefont {{Bharadwaj}}\ and\ \citenamefont
  {{Sethi}}(2001)}]{poreion1}%
  \BibitemOpen
  \bibfield  {author} {\bibinfo {author} {\bibfnamefont {S.}~\bibnamefont
  {{Bharadwaj}}}\ and\ \bibinfo {author} {\bibfnamefont {S.~K.}\ \bibnamefont
  {{Sethi}}},\ }\href@noop {} {\bibfield  {journal} {\bibinfo  {journal}
  {Journal of Astrophysics and Astronomy}\ }\textbf {\bibinfo {volume} {22}},\
  \bibinfo {pages} {293} (\bibinfo {year} {2001})},\ \Eprint
  {https://arxiv.org/abs/arXiv:astro-ph/0203269} {arXiv:astro-ph/0203269}
  \BibitemShut {NoStop}%
\bibitem [{\citenamefont {{Bharadwaj}}\ \emph {et~al.}(2001)\citenamefont
  {{Bharadwaj}}, \citenamefont {{Nath}},\ and\ \citenamefont
  {{Sethi}}}]{poreion2}%
  \BibitemOpen
  \bibfield  {author} {\bibinfo {author} {\bibfnamefont {S.}~\bibnamefont
  {{Bharadwaj}}}, \bibinfo {author} {\bibfnamefont {B.~B.}\ \bibnamefont
  {{Nath}}},\ and\ \bibinfo {author} {\bibfnamefont {S.~K.}\ \bibnamefont
  {{Sethi}}},\ }\href@noop {} {\bibfield  {journal} {\bibinfo  {journal}
  {Journal of Astrophysics and Astronomy}\ }\textbf {\bibinfo {volume} {22}},\
  \bibinfo {pages} {21} (\bibinfo {year} {2001})},\ \Eprint
  {https://arxiv.org/abs/arXiv:astro-ph/0003200} {arXiv:astro-ph/0003200}
  \BibitemShut {NoStop}%
\bibitem [{\citenamefont {{Wyithe}}\ and\ \citenamefont
  {{Loeb}}(2007)}]{poreion3}%
  \BibitemOpen
  \bibfield  {author} {\bibinfo {author} {\bibfnamefont {S.}~\bibnamefont
  {{Wyithe}}}\ and\ \bibinfo {author} {\bibfnamefont {A.}~\bibnamefont
  {{Loeb}}},\ }\href@noop {} {\bibfield  {journal} {\bibinfo  {journal} {ArXiv
  e-prints}\ } (\bibinfo {year} {2007})},\ \Eprint
  {https://arxiv.org/abs/0708.3392} {arXiv:0708.3392} \BibitemShut {NoStop}%
\bibitem [{\citenamefont {{Loeb}}\ and\ \citenamefont
  {{Wyithe}}(2008)}]{poreion4}%
  \BibitemOpen
  \bibfield  {author} {\bibinfo {author} {\bibfnamefont {A.}~\bibnamefont
  {{Loeb}}}\ and\ \bibinfo {author} {\bibfnamefont {J.~S.~B.}\ \bibnamefont
  {{Wyithe}}},\ }\href@noop {} {\bibfield  {journal} {\bibinfo  {journal}
  {Physical Review Letters}\ }\textbf {\bibinfo {volume} {100}},\ \bibinfo
  {pages} {161301} (\bibinfo {year} {2008})},\ \Eprint
  {https://arxiv.org/abs/0801.1677} {arXiv:0801.1677} \BibitemShut {NoStop}%
\bibitem [{\citenamefont {{Wyithe}}\ and\ \citenamefont
  {{Loeb}}(2008)}]{poreion5}%
  \BibitemOpen
  \bibfield  {author} {\bibinfo {author} {\bibfnamefont {S.}~\bibnamefont
  {{Wyithe}}}\ and\ \bibinfo {author} {\bibfnamefont {A.}~\bibnamefont
  {{Loeb}}},\ }\href@noop {} {\bibfield  {journal} {\bibinfo  {journal} {ArXiv
  e-prints}\ } (\bibinfo {year} {2008})},\ \Eprint
  {https://arxiv.org/abs/0808.2323} {arXiv:0808.2323} \BibitemShut {NoStop}%
\bibitem [{\citenamefont {{Visbal}}\ \emph {et~al.}(2009)\citenamefont
  {{Visbal}}, \citenamefont {{Loeb}},\ and\ \citenamefont
  {{Wyithe}}}]{poreion6}%
  \BibitemOpen
  \bibfield  {author} {\bibinfo {author} {\bibfnamefont {E.}~\bibnamefont
  {{Visbal}}}, \bibinfo {author} {\bibfnamefont {A.}~\bibnamefont {{Loeb}}},\
  and\ \bibinfo {author} {\bibfnamefont {S.}~\bibnamefont {{Wyithe}}},\
  }\href@noop {} {\bibfield  {journal} {\bibinfo  {journal} {Journal of
  Cosmology and Astro-Particle Physics}\ }\textbf {\bibinfo {volume} {10}},\
  \bibinfo {pages} {30} (\bibinfo {year} {2009})},\ \Eprint
  {https://arxiv.org/abs/0812.0419} {arXiv:0812.0419} \BibitemShut {NoStop}%
\bibitem [{\citenamefont {{Bharadwaj}}\ and\ \citenamefont
  {{Pandey}}(2003)}]{poreion7}%
  \BibitemOpen
  \bibfield  {author} {\bibinfo {author} {\bibfnamefont {S.}~\bibnamefont
  {{Bharadwaj}}}\ and\ \bibinfo {author} {\bibfnamefont {S.~K.}\ \bibnamefont
  {{Pandey}}},\ }\href@noop {} {\bibfield  {journal} {\bibinfo  {journal}
  {Journal of Astrophysics and Astronomy}\ }\textbf {\bibinfo {volume} {24}},\
  \bibinfo {pages} {23} (\bibinfo {year} {2003})},\ \Eprint
  {https://arxiv.org/abs/arXiv:astro-ph/0307303} {arXiv:astro-ph/0307303}
  \BibitemShut {NoStop}%
\bibitem [{\citenamefont {{Bharadwaj}}\ and\ \citenamefont
  {{Srikant}}(2004)}]{poreion8}%
  \BibitemOpen
  \bibfield  {author} {\bibinfo {author} {\bibfnamefont {S.}~\bibnamefont
  {{Bharadwaj}}}\ and\ \bibinfo {author} {\bibfnamefont {P.~S.}\ \bibnamefont
  {{Srikant}}},\ }\href {https://doi.org/10.1007/BF02702289} {\bibfield
  {journal} {\bibinfo  {journal} {Journal of Astrophysics and Astronomy}\
  }\textbf {\bibinfo {volume} {25}},\ \bibinfo {pages} {67} (\bibinfo {year}
  {2004})},\ \Eprint {https://arxiv.org/abs/arXiv:astro-ph/0402262}
  {arXiv:astro-ph/0402262} \BibitemShut {NoStop}%
\bibitem [{\citenamefont {{Subramanian}}\ and\ \citenamefont
  {{Padmanabhan}}(1993)}]{poreion9}%
  \BibitemOpen
  \bibfield  {author} {\bibinfo {author} {\bibfnamefont {K.}~\bibnamefont
  {{Subramanian}}}\ and\ \bibinfo {author} {\bibfnamefont {T.}~\bibnamefont
  {{Padmanabhan}}},\ }\href {https://doi.org/10.1093/mnras/265.1.101}
  {\bibfield  {journal} {\bibinfo  {journal} {MNRAS}\ }\textbf {\bibinfo
  {volume} {265}},\ \bibinfo {pages} {101} (\bibinfo {year}
  {1993})}\BibitemShut {NoStop}%
\bibitem [{\citenamefont {{Kumar}}\ \emph {et~al.}(1995)\citenamefont
  {{Kumar}}, \citenamefont {{Padmanabhan}},\ and\ \citenamefont
  {{Subramanian}}}]{poreion10}%
  \BibitemOpen
  \bibfield  {author} {\bibinfo {author} {\bibfnamefont {A.}~\bibnamefont
  {{Kumar}}}, \bibinfo {author} {\bibfnamefont {T.}~\bibnamefont
  {{Padmanabhan}}},\ and\ \bibinfo {author} {\bibfnamefont {K.}~\bibnamefont
  {{Subramanian}}},\ }\href {https://doi.org/10.1093/mnras/272.3.544}
  {\bibfield  {journal} {\bibinfo  {journal} {MNRAS}\ }\textbf {\bibinfo
  {volume} {272}},\ \bibinfo {pages} {544} (\bibinfo {year}
  {1995})}\BibitemShut {NoStop}%
\bibitem [{\citenamefont {{Bagla}}\ \emph {et~al.}(1997)\citenamefont
  {{Bagla}}, \citenamefont {{Nath}},\ and\ \citenamefont
  {{Padmanabhan}}}]{poreion11}%
  \BibitemOpen
  \bibfield  {author} {\bibinfo {author} {\bibfnamefont {J.~S.}\ \bibnamefont
  {{Bagla}}}, \bibinfo {author} {\bibfnamefont {B.}~\bibnamefont {{Nath}}},\
  and\ \bibinfo {author} {\bibfnamefont {T.}~\bibnamefont {{Padmanabhan}}},\
  }\href@noop {} {\bibfield  {journal} {\bibinfo  {journal} {MNRAS}\ }\textbf
  {\bibinfo {volume} {289}},\ \bibinfo {pages} {671} (\bibinfo {year}
  {1997})},\ \Eprint {https://arxiv.org/abs/arXiv:astro-ph/9610267}
  {arXiv:astro-ph/9610267} \BibitemShut {NoStop}%
\bibitem [{\citenamefont {{Padmanabhan}}\ \emph {et~al.}(2015)\citenamefont
  {{Padmanabhan}}, \citenamefont {Choudhury},\ and\ \citenamefont
  {Refregier}}]{poreion12}%
  \BibitemOpen
  \bibfield  {author} {\bibinfo {author} {\bibfnamefont {H.}~\bibnamefont
  {{Padmanabhan}}}, \bibinfo {author} {\bibfnamefont {T.~R.}\ \bibnamefont
  {Choudhury}},\ and\ \bibinfo {author} {\bibfnamefont {A.}~\bibnamefont
  {Refregier}},\ }\href {https://doi.org/10.1093/mnras/stu2702} {\bibfield
  {journal} {\bibinfo  {journal} {Monthly Notices of the Royal Astronomical
  Society}\ }\textbf {\bibinfo {volume} {447}},\ \bibinfo {pages} {3745}
  (\bibinfo {year} {2015})}\BibitemShut {NoStop}%
\bibitem [{\citenamefont {Bagla}\ \emph {et~al.}(2010)\citenamefont {Bagla},
  \citenamefont {Khandai},\ and\ \citenamefont {Datta}}]{Bagla_2010}%
  \BibitemOpen
  \bibfield  {author} {\bibinfo {author} {\bibfnamefont {J.~S.}\ \bibnamefont
  {Bagla}}, \bibinfo {author} {\bibfnamefont {N.}~\bibnamefont {Khandai}},\
  and\ \bibinfo {author} {\bibfnamefont {K.~K.}\ \bibnamefont {Datta}},\ }\href
  {https://doi.org/10.1111/j.1365-2966.2010.16933.x} {\bibfield  {journal}
  {\bibinfo  {journal} {Monthly Notices of the Royal Astronomical Society}\
  }\textbf {\bibinfo {volume} {407}},\ \bibinfo {pages} {567–580} (\bibinfo
  {year} {2010})}\BibitemShut {NoStop}%
\bibitem [{\citenamefont {Guha~Sarkar}\ \emph {et~al.}(2012)\citenamefont
  {Guha~Sarkar}, \citenamefont {Mitra}, \citenamefont {Majumdar},\ and\
  \citenamefont {Choudhury}}]{Guha_Sarkar_2012}%
  \BibitemOpen
  \bibfield  {author} {\bibinfo {author} {\bibfnamefont {T.}~\bibnamefont
  {Guha~Sarkar}}, \bibinfo {author} {\bibfnamefont {S.}~\bibnamefont {Mitra}},
  \bibinfo {author} {\bibfnamefont {S.}~\bibnamefont {Majumdar}},\ and\
  \bibinfo {author} {\bibfnamefont {T.~R.}\ \bibnamefont {Choudhury}},\ }\href
  {https://doi.org/10.1111/j.1365-2966.2012.20582.x} {\bibfield  {journal}
  {\bibinfo  {journal} {Monthly Notices of the Royal Astronomical Society}\
  }\textbf {\bibinfo {volume} {421}},\ \bibinfo {pages} {3570–3578} (\bibinfo
  {year} {2012})}\BibitemShut {NoStop}%
\bibitem [{\citenamefont {Sarkar}\ \emph {et~al.}(2016)\citenamefont {Sarkar},
  \citenamefont {Bharadwaj},\ and\ \citenamefont {Anathpindika}}]{Sarkar_2016}%
  \BibitemOpen
  \bibfield  {author} {\bibinfo {author} {\bibfnamefont {D.}~\bibnamefont
  {Sarkar}}, \bibinfo {author} {\bibfnamefont {S.}~\bibnamefont {Bharadwaj}},\
  and\ \bibinfo {author} {\bibfnamefont {S.}~\bibnamefont {Anathpindika}},\
  }\href {https://doi.org/10.1093/mnras/stw1111} {\bibfield  {journal}
  {\bibinfo  {journal} {Monthly Notices of the Royal Astronomical Society}\
  }\textbf {\bibinfo {volume} {460}},\ \bibinfo {pages} {4310–4319} (\bibinfo
  {year} {2016})}\BibitemShut {NoStop}%
\bibitem [{\citenamefont {{Storrie-Lombardi}}\ \emph
  {et~al.}(1996)\citenamefont {{Storrie-Lombardi}}, \citenamefont {{McMahon}},\
  and\ \citenamefont {{Irwin}}}]{xhibar1}%
  \BibitemOpen
  \bibfield  {author} {\bibinfo {author} {\bibfnamefont {L.~J.}\ \bibnamefont
  {{Storrie-Lombardi}}}, \bibinfo {author} {\bibfnamefont {R.~G.}\ \bibnamefont
  {{McMahon}}},\ and\ \bibinfo {author} {\bibfnamefont {M.~J.}\ \bibnamefont
  {{Irwin}}},\ }\href@noop {} {\bibfield  {journal} {\bibinfo  {journal}
  {MNRAS}\ }\textbf {\bibinfo {volume} {283}},\ \bibinfo {pages} {L79}
  (\bibinfo {year} {1996})},\ \Eprint
  {https://arxiv.org/abs/arXiv:astro-ph/9608147} {arXiv:astro-ph/9608147}
  \BibitemShut {NoStop}%
\bibitem [{\citenamefont {{Peroux}}\ \emph {et~al.}(2003)\citenamefont
  {{Peroux}}, \citenamefont {{McMahon}}, \citenamefont {{Storrie-Lombardi}},\
  and\ \citenamefont {{Irwin}}}]{xhibar2}%
  \BibitemOpen
  \bibfield  {author} {\bibinfo {author} {\bibfnamefont {C.}~\bibnamefont
  {{Peroux}}}, \bibinfo {author} {\bibfnamefont {R.~G.}\ \bibnamefont
  {{McMahon}}}, \bibinfo {author} {\bibfnamefont {L.~J.}\ \bibnamefont
  {{Storrie-Lombardi}}},\ and\ \bibinfo {author} {\bibfnamefont {M.~J.}\
  \bibnamefont {{Irwin}}},\ }\href@noop {} {\bibfield  {journal} {\bibinfo
  {journal} {MNRAS}\ }\textbf {\bibinfo {volume} {346}},\ \bibinfo {pages}
  {1103} (\bibinfo {year} {2003})},\ \Eprint
  {https://arxiv.org/abs/arXiv:astro-ph/0107045} {arXiv:astro-ph/0107045}
  \BibitemShut {NoStop}%
\bibitem [{\citenamefont {McQuinn}\ \emph {et~al.}(2006)\citenamefont
  {McQuinn}, \citenamefont {Zahn}, \citenamefont {Zaldarriaga}, \citenamefont
  {Hernquist},\ and\ \citenamefont {Furlanetto}}]{mcquinn2006cosmological}%
  \BibitemOpen
  \bibfield  {author} {\bibinfo {author} {\bibfnamefont {M.}~\bibnamefont
  {McQuinn}}, \bibinfo {author} {\bibfnamefont {O.}~\bibnamefont {Zahn}},
  \bibinfo {author} {\bibfnamefont {M.}~\bibnamefont {Zaldarriaga}}, \bibinfo
  {author} {\bibfnamefont {L.}~\bibnamefont {Hernquist}},\ and\ \bibinfo
  {author} {\bibfnamefont {S.~R.}\ \bibnamefont {Furlanetto}},\ }\href
  {https://doi.org/10.1086/505167} {\bibfield  {journal} {\bibinfo  {journal}
  {The Astrophysical Journal}\ }\textbf {\bibinfo {volume} {653}},\ \bibinfo
  {pages} {815} (\bibinfo {year} {2006})}\BibitemShut {NoStop}%
\bibitem [{\citenamefont {Bull}\ \emph {et~al.}(2015)\citenamefont {Bull},
  \citenamefont {Ferreira}, \citenamefont {Patel},\ and\ \citenamefont
  {Santos}}]{Bull_2015}%
  \BibitemOpen
  \bibfield  {author} {\bibinfo {author} {\bibfnamefont {P.}~\bibnamefont
  {Bull}}, \bibinfo {author} {\bibfnamefont {P.~G.}\ \bibnamefont {Ferreira}},
  \bibinfo {author} {\bibfnamefont {P.}~\bibnamefont {Patel}},\ and\ \bibinfo
  {author} {\bibfnamefont {M.~G.}\ \bibnamefont {Santos}},\ }\href
  {https://doi.org/10.1088/0004-637x/803/1/21} {\bibfield  {journal} {\bibinfo
  {journal} {The Astrophysical Journal}\ }\textbf {\bibinfo {volume} {803}},\
  \bibinfo {pages} {21} (\bibinfo {year} {2015})}\BibitemShut {NoStop}%
\bibitem [{\citenamefont {Chavan}\ \emph
  {et~al.}(2025{\natexlab{a}})\citenamefont {Chavan}, \citenamefont {Sarkar},\
  and\ \citenamefont {Sen}}]{P_chavan_2025_semiCosmography_xpf8}%
  \BibitemOpen
  \bibfield  {author} {\bibinfo {author} {\bibfnamefont {P.}~\bibnamefont
  {Chavan}}, \bibinfo {author} {\bibfnamefont {T.~G.}\ \bibnamefont {Sarkar}},\
  and\ \bibinfo {author} {\bibfnamefont {A.~A.}\ \bibnamefont {Sen}},\ }\href
  {https://arxiv.org/abs/2506.14275} {\bibinfo {title} {The dynamics of
  background evolution and structure formation in phase space: a
  semi-cosmographic reconstruction}} (\bibinfo {year} {2025}{\natexlab{a}}),\
  \Eprint {https://arxiv.org/abs/2506.14275} {arXiv:2506.14275 [astro-ph.CO]}
  \BibitemShut {NoStop}%
\bibitem [{\citenamefont {Foreman-Mackey}\ \emph {et~al.}(2013)\citenamefont
  {Foreman-Mackey}, \citenamefont {Hogg}, \citenamefont {Lang},\ and\
  \citenamefont {Goodman}}]{foreman2013emcee}%
  \BibitemOpen
  \bibfield  {author} {\bibinfo {author} {\bibfnamefont {D.}~\bibnamefont
  {Foreman-Mackey}}, \bibinfo {author} {\bibfnamefont {D.~W.}\ \bibnamefont
  {Hogg}}, \bibinfo {author} {\bibfnamefont {D.}~\bibnamefont {Lang}},\ and\
  \bibinfo {author} {\bibfnamefont {J.}~\bibnamefont {Goodman}},\ }\href
  {https://doi.org/10.1086/670067} {\bibfield  {journal} {\bibinfo  {journal}
  {Publications of the Astronomical Society of the Pacific}\ }\textbf {\bibinfo
  {volume} {125}},\ \bibinfo {pages} {306} (\bibinfo {year}
  {2013})}\BibitemShut {NoStop}%
\bibitem [{\citenamefont {Kazantzidis}\ and\ \citenamefont
  {Perivolaropoulos}(2018)}]{Kazantzidis_Perivolaropoulos_fs8_2018}%
  \BibitemOpen
  \bibfield  {author} {\bibinfo {author} {\bibfnamefont {L.}~\bibnamefont
  {Kazantzidis}}\ and\ \bibinfo {author} {\bibfnamefont {L.}~\bibnamefont
  {Perivolaropoulos}},\ }\href {https://doi.org/10.1103/PhysRevD.97.103503}
  {\bibfield  {journal} {\bibinfo  {journal} {Phys. Rev. D}\ }\textbf {\bibinfo
  {volume} {97}},\ \bibinfo {pages} {103503} (\bibinfo {year}
  {2018})}\BibitemShut {NoStop}%
\bibitem [{\citenamefont {Nesseris}\ and\ \citenamefont
  {García-Bellido}(2012)}]{Savvas_Nesseris_2012_fs8_SN_BAO}%
  \BibitemOpen
  \bibfield  {author} {\bibinfo {author} {\bibfnamefont {S.}~\bibnamefont
  {Nesseris}}\ and\ \bibinfo {author} {\bibfnamefont {J.}~\bibnamefont
  {García-Bellido}},\ }\href {https://doi.org/10.1088/1475-7516/2012/11/033}
  {\bibfield  {journal} {\bibinfo  {journal} {Journal of Cosmology and
  Astroparticle Physics}\ }\textbf {\bibinfo {volume} {2012}}\bibinfo  {number}
  { (11)},\ \bibinfo {pages} {033}}\BibitemShut {NoStop}%
\bibitem [{\citenamefont {Chavan}\ \emph
  {et~al.}(2025{\natexlab{b}})\citenamefont {Chavan}, \citenamefont {Sarkar},
  \citenamefont {Dash},\ and\ \citenamefont
  {Sen}}]{P_Chavan_2025_semiCosmography}%
  \BibitemOpen
\bibfield  {number} {  }\bibfield  {author} {\bibinfo {author} {\bibfnamefont
  {P.}~\bibnamefont {Chavan}}, \bibinfo {author} {\bibfnamefont {T.~G.}\
  \bibnamefont {Sarkar}}, \bibinfo {author} {\bibfnamefont {C.~B.~V.}\
  \bibnamefont {Dash}},\ and\ \bibinfo {author} {\bibfnamefont {A.~A.}\
  \bibnamefont {Sen}},\ }\href {https://arxiv.org/abs/2503.03288} {\bibinfo
  {title} {A semi-cosmographic approach to study cosmological evolution in
  phase space}} (\bibinfo {year} {2025}{\natexlab{b}}),\ \Eprint
  {https://arxiv.org/abs/2503.03288} {arXiv:2503.03288 [astro-ph.CO]}
  \BibitemShut {NoStop}%
\bibitem [{\citenamefont {Saini}\ \emph {et~al.}(2000)\citenamefont {Saini},
  \citenamefont {Raychaudhury}, \citenamefont {Sahni},\ and\ \citenamefont
  {Starobinsky}}]{Saini_2000}%
  \BibitemOpen
  \bibfield  {author} {\bibinfo {author} {\bibfnamefont {T.~D.}\ \bibnamefont
  {Saini}}, \bibinfo {author} {\bibfnamefont {S.}~\bibnamefont {Raychaudhury}},
  \bibinfo {author} {\bibfnamefont {V.}~\bibnamefont {Sahni}},\ and\ \bibinfo
  {author} {\bibfnamefont {A.~A.}\ \bibnamefont {Starobinsky}},\ }\href
  {https://doi.org/10.1103/physrevlett.85.1162} {\bibfield  {journal} {\bibinfo
   {journal} {Physical Review Letters}\ }\textbf {\bibinfo {volume} {85}},\
  \bibinfo {pages} {1162–1165} (\bibinfo {year} {2000})}\BibitemShut
  {NoStop}%
\bibitem [{\citenamefont {Holsclaw}\ \emph {et~al.}(2011)\citenamefont
  {Holsclaw}, \citenamefont {Alam}, \citenamefont {Sans\'o}, \citenamefont
  {Lee}, \citenamefont {Heitmann}, \citenamefont {Habib},\ and\ \citenamefont
  {Higdon}}]{Holsclaw_2011_GPR}%
  \BibitemOpen
  \bibfield  {author} {\bibinfo {author} {\bibfnamefont {T.}~\bibnamefont
  {Holsclaw}}, \bibinfo {author} {\bibfnamefont {U.}~\bibnamefont {Alam}},
  \bibinfo {author} {\bibfnamefont {B.}~\bibnamefont {Sans\'o}}, \bibinfo
  {author} {\bibfnamefont {H.}~\bibnamefont {Lee}}, \bibinfo {author}
  {\bibfnamefont {K.}~\bibnamefont {Heitmann}}, \bibinfo {author}
  {\bibfnamefont {S.}~\bibnamefont {Habib}},\ and\ \bibinfo {author}
  {\bibfnamefont {D.}~\bibnamefont {Higdon}},\ }\href
  {https://doi.org/10.1103/PhysRevD.84.083501} {\bibfield  {journal} {\bibinfo
  {journal} {Phys. Rev. D}\ }\textbf {\bibinfo {volume} {84}},\ \bibinfo
  {pages} {083501} (\bibinfo {year} {2011})}\BibitemShut {NoStop}%
\bibitem [{\citenamefont {Shafieloo}\ \emph {et~al.}(2012)\citenamefont
  {Shafieloo}, \citenamefont {Kim},\ and\ \citenamefont
  {Linder}}]{Shafieloo_2012_GPR}%
  \BibitemOpen
  \bibfield  {author} {\bibinfo {author} {\bibfnamefont {A.}~\bibnamefont
  {Shafieloo}}, \bibinfo {author} {\bibfnamefont {A.~G.}\ \bibnamefont {Kim}},\
  and\ \bibinfo {author} {\bibfnamefont {E.~V.}\ \bibnamefont {Linder}},\
  }\href {https://doi.org/10.1103/PhysRevD.85.123530} {\bibfield  {journal}
  {\bibinfo  {journal} {Phys. Rev. D}\ }\textbf {\bibinfo {volume} {85}},\
  \bibinfo {pages} {123530} (\bibinfo {year} {2012})}\BibitemShut {NoStop}%
\bibitem [{\citenamefont {Jesus}\ \emph {et~al.}(2024)\citenamefont {Jesus},
  \citenamefont {Benndorf}, \citenamefont {Escobal},\ and\ \citenamefont
  {Pereira}}]{Jesus_2024_GPR}%
  \BibitemOpen
  \bibfield  {author} {\bibinfo {author} {\bibfnamefont {J.~F.}\ \bibnamefont
  {Jesus}}, \bibinfo {author} {\bibfnamefont {D.}~\bibnamefont {Benndorf}},
  \bibinfo {author} {\bibfnamefont {A.~A.}\ \bibnamefont {Escobal}},\ and\
  \bibinfo {author} {\bibfnamefont {S.~H.}\ \bibnamefont {Pereira}},\ }\href
  {https://doi.org/10.1093/mnras/stae120} {\bibfield  {journal} {\bibinfo
  {journal} {Monthly Notices of the Royal Astronomical Society}\ }\textbf
  {\bibinfo {volume} {528}},\ \bibinfo {pages} {1573} (\bibinfo {year}
  {2024})},\ \Eprint
  {https://arxiv.org/abs/https://academic.oup.com/mnras/article-pdf/528/2/1573/56410686/stae120.pdf}
  {https://academic.oup.com/mnras/article-pdf/528/2/1573/56410686/stae120.pdf}
  \BibitemShut {NoStop}%
\bibitem [{\citenamefont {Dinda}(2024)}]{Dinda2024_GP_cosmography}%
  \BibitemOpen
  \bibfield  {author} {\bibinfo {author} {\bibfnamefont {B.~R.}\ \bibnamefont
  {Dinda}},\ }\href {https://doi.org/10.1140/epjc/s10052-024-12774-x}
  {\bibfield  {journal} {\bibinfo  {journal} {The European Physical Journal C}\
  }\textbf {\bibinfo {volume} {84}},\ \bibinfo {pages} {402} (\bibinfo {year}
  {2024})}\BibitemShut {NoStop}%
\bibitem [{\citenamefont {Velázquez}\ \emph {et~al.}(2024)\citenamefont
  {Velázquez}, \citenamefont {Escamilla}, \citenamefont {Mukherjee},\ and\
  \citenamefont {Vázquez}}]{Velazquez_mukherjee_GPR_2024}%
  \BibitemOpen
  \bibfield  {author} {\bibinfo {author} {\bibfnamefont {J.~d.~J.}\
  \bibnamefont {Velázquez}}, \bibinfo {author} {\bibfnamefont {L.~A.}\
  \bibnamefont {Escamilla}}, \bibinfo {author} {\bibfnamefont {P.}~\bibnamefont
  {Mukherjee}},\ and\ \bibinfo {author} {\bibfnamefont {J.~A.}\ \bibnamefont
  {Vázquez}},\ }\bibfield  {journal} {\bibinfo  {journal} {Universe}\ }\textbf
  {\bibinfo {volume} {10}},\ \href {https://doi.org/10.3390/universe10120464}
  {10.3390/universe10120464} (\bibinfo {year} {2024})\BibitemShut {NoStop}%
\bibitem [{\citenamefont {Mukherjee}\ and\ \citenamefont
  {Sen}(2024)}]{Purba_Mukherjee-Anjan_sen_GPR_2024}%
  \BibitemOpen
  \bibfield  {author} {\bibinfo {author} {\bibfnamefont {P.}~\bibnamefont
  {Mukherjee}}\ and\ \bibinfo {author} {\bibfnamefont {A.~A.}\ \bibnamefont
  {Sen}},\ }\href {https://doi.org/10.1103/PhysRevD.110.123502} {\bibfield
  {journal} {\bibinfo  {journal} {Phys. Rev. D}\ }\textbf {\bibinfo {volume}
  {110}},\ \bibinfo {pages} {123502} (\bibinfo {year} {2024})}\BibitemShut
  {NoStop}%
\bibitem [{\citenamefont {Dinda}\ and\ \citenamefont
  {Maartens}(2025)}]{Dinda_2025}%
  \BibitemOpen
  \bibfield  {author} {\bibinfo {author} {\bibfnamefont {B.~R.}\ \bibnamefont
  {Dinda}}\ and\ \bibinfo {author} {\bibfnamefont {R.}~\bibnamefont
  {Maartens}},\ }\href {https://doi.org/10.1088/1475-7516/2025/01/120}
  {\bibfield  {journal} {\bibinfo  {journal} {Journal of Cosmology and
  Astroparticle Physics}\ }\textbf {\bibinfo {volume} {2025}}\bibinfo  {number}
  { (01)},\ \bibinfo {pages} {120}}\BibitemShut {NoStop}%
\bibitem [{\citenamefont {{Bharadwaj}}\ and\ \citenamefont
  {{Ali}}(2005)}]{bali}%
  \BibitemOpen
\bibfield  {number} {  }\bibfield  {author} {\bibinfo {author} {\bibfnamefont
  {S.}~\bibnamefont {{Bharadwaj}}}\ and\ \bibinfo {author} {\bibfnamefont
  {S.~S.}\ \bibnamefont {{Ali}}},\ }\href@noop {} {\bibfield  {journal}
  {\bibinfo  {journal} {MNRAS}\ }\textbf {\bibinfo {volume} {356}},\ \bibinfo
  {pages} {1519} (\bibinfo {year} {2005})},\ \Eprint
  {https://arxiv.org/abs/arXiv:astro-ph/0406676} {arXiv:astro-ph/0406676}
  \BibitemShut {NoStop}%
\bibitem [{\citenamefont {Sarkar}\ \emph {et~al.}(2017)\citenamefont {Sarkar},
  \citenamefont {Bharadwaj},\ and\ \citenamefont {Marthi}}]{Sarkar_2017}%
  \BibitemOpen
  \bibfield  {author} {\bibinfo {author} {\bibfnamefont {A.~K.}\ \bibnamefont
  {Sarkar}}, \bibinfo {author} {\bibfnamefont {S.}~\bibnamefont {Bharadwaj}},\
  and\ \bibinfo {author} {\bibfnamefont {V.~R.}\ \bibnamefont {Marthi}},\
  }\href {https://doi.org/10.1093/mnras/stx2344} {\bibfield  {journal}
  {\bibinfo  {journal} {Monthly Notices of the Royal Astronomical Society}\
  }\textbf {\bibinfo {volume} {473}},\ \bibinfo {pages} {261–270} (\bibinfo
  {year} {2017})}\BibitemShut {NoStop}%
\bibitem [{Note1()}]{Note1}%
  \BibitemOpen
  \bibinfo {note} {Https://www.skao.int/en}\BibitemShut {NoStop}%
\bibitem [{\citenamefont {Di~Matteo}\ \emph {et~al.}(2002)\citenamefont
  {Di~Matteo}, \citenamefont {Perna}, \citenamefont {Abel},\ and\ \citenamefont
  {Rees}}]{di2002radio}%
  \BibitemOpen
  \bibfield  {author} {\bibinfo {author} {\bibfnamefont {T.}~\bibnamefont
  {Di~Matteo}}, \bibinfo {author} {\bibfnamefont {R.}~\bibnamefont {Perna}},
  \bibinfo {author} {\bibfnamefont {T.}~\bibnamefont {Abel}},\ and\ \bibinfo
  {author} {\bibfnamefont {M.~J.}\ \bibnamefont {Rees}},\ }\href
  {https://doi.org/10.1086/324293} {\bibfield  {journal} {\bibinfo  {journal}
  {The Astrophysical Journal}\ }\textbf {\bibinfo {volume} {564}},\ \bibinfo
  {pages} {576} (\bibinfo {year} {2002})}\BibitemShut {NoStop}%
\bibitem [{\citenamefont {Ghosh}\ \emph {et~al.}(2010)\citenamefont {Ghosh},
  \citenamefont {Bharadwaj}, \citenamefont {Ali},\ and\ \citenamefont
  {Chengalur}}]{Ghosh_2010}%
  \BibitemOpen
  \bibfield  {author} {\bibinfo {author} {\bibfnamefont {A.}~\bibnamefont
  {Ghosh}}, \bibinfo {author} {\bibfnamefont {S.}~\bibnamefont {Bharadwaj}},
  \bibinfo {author} {\bibfnamefont {S.~S.}\ \bibnamefont {Ali}},\ and\ \bibinfo
  {author} {\bibfnamefont {J.~N.}\ \bibnamefont {Chengalur}},\ }\href
  {https://doi.org/10.1111/j.1365-2966.2010.17853.x} {\bibfield  {journal}
  {\bibinfo  {journal} {Monthly Notices of the Royal Astronomical Society}\
  }\textbf {\bibinfo {volume} {411}},\ \bibinfo {pages} {2426–2438} (\bibinfo
  {year} {2010})}\BibitemShut {NoStop}%
\bibitem [{\citenamefont {Wang}\ \emph {et~al.}(2006)\citenamefont {Wang},
  \citenamefont {Tegmark}, \citenamefont {Santos},\ and\ \citenamefont
  {Knox}}]{wang200621}%
  \BibitemOpen
  \bibfield  {author} {\bibinfo {author} {\bibfnamefont {X.}~\bibnamefont
  {Wang}}, \bibinfo {author} {\bibfnamefont {M.}~\bibnamefont {Tegmark}},
  \bibinfo {author} {\bibfnamefont {M.~G.}\ \bibnamefont {Santos}},\ and\
  \bibinfo {author} {\bibfnamefont {L.}~\bibnamefont {Knox}},\ }\href
  {https://doi.org/10.1086/506597} {\bibfield  {journal} {\bibinfo  {journal}
  {The Astrophysical Journal}\ }\textbf {\bibinfo {volume} {650}},\ \bibinfo
  {pages} {529} (\bibinfo {year} {2006})}\BibitemShut {NoStop}%
\bibitem [{\citenamefont {{Liu}}\ \emph {et~al.}(2014)\citenamefont {{Liu}},
  \citenamefont {{Parsons}},\ and\ \citenamefont {{Trott}}}]{Liu-formalism1}%
  \BibitemOpen
  \bibfield  {author} {\bibinfo {author} {\bibfnamefont {A.}~\bibnamefont
  {{Liu}}}, \bibinfo {author} {\bibfnamefont {A.~R.}\ \bibnamefont
  {{Parsons}}},\ and\ \bibinfo {author} {\bibfnamefont {C.~M.}\ \bibnamefont
  {{Trott}}},\ }\href {https://doi.org/10.1103/PhysRevD.90.023018} {\bibfield
  {journal} {\bibinfo  {journal} {\prd}\ }\textbf {\bibinfo {volume} {90}},\
  \bibinfo {eid} {023018} (\bibinfo {year} {2014})},\ \Eprint
  {https://arxiv.org/abs/1404.2596} {arXiv:1404.2596 [astro-ph.CO]}
  \BibitemShut {NoStop}%
\bibitem [{\citenamefont {Liu}\ and\ \citenamefont
  {Tegmark}(2012)}]{liu2012well}%
  \BibitemOpen
  \bibfield  {author} {\bibinfo {author} {\bibfnamefont {A.}~\bibnamefont
  {Liu}}\ and\ \bibinfo {author} {\bibfnamefont {M.}~\bibnamefont {Tegmark}},\
  }\href {https://doi.org/10.1111/j.1365-2966.2011.19989.x} {\bibfield
  {journal} {\bibinfo  {journal} {Monthly Notices of the Royal Astronomical
  Society}\ }\textbf {\bibinfo {volume} {419}},\ \bibinfo {pages} {3491}
  (\bibinfo {year} {2012})}\BibitemShut {NoStop}%
\bibitem [{\citenamefont {Liu}\ \emph {et~al.}(2009)\citenamefont {Liu},
  \citenamefont {Tegmark}, \citenamefont {Bowman}, \citenamefont {Hewitt},\
  and\ \citenamefont {Zaldarriaga}}]{liu2009improved}%
  \BibitemOpen
  \bibfield  {author} {\bibinfo {author} {\bibfnamefont {A.}~\bibnamefont
  {Liu}}, \bibinfo {author} {\bibfnamefont {M.}~\bibnamefont {Tegmark}},
  \bibinfo {author} {\bibfnamefont {J.}~\bibnamefont {Bowman}}, \bibinfo
  {author} {\bibfnamefont {J.}~\bibnamefont {Hewitt}},\ and\ \bibinfo {author}
  {\bibfnamefont {M.}~\bibnamefont {Zaldarriaga}},\ }\href
  {https://doi.org/10.1111/j.1365-2966.2009.15156.x} {\bibfield  {journal}
  {\bibinfo  {journal} {Monthly Notices of the Royal Astronomical Society}\
  }\textbf {\bibinfo {volume} {398}},\ \bibinfo {pages} {401} (\bibinfo {year}
  {2009})}\BibitemShut {NoStop}%
\end{thebibliography}%
\end{document}